\documentstyle[12pt]{article}

\input epsf.sty

\newcommand{\g}{{\sl g}}
\newcommand{\BRST}{{\rm BRST}}
\newcommand{\J}{{\cal J}}
\newcommand{\X}{{\cal X}}
\newcommand{\D}{{\cal D}}
\newcommand{\K}{{\cal K}}

\newcommand{\W}{{\cal W}}
\newcommand{\V}{{\cal V}}
\newcommand{\I}{{\cal I}}

\begin{document}
\begin{titlepage}

\centerline{\large \bf Broken conformal invariance and spectrum }
\centerline{\large \bf of anomalous dimensions in QCD.}

\vspace{10mm}

\centerline{\bf A.V. Belitsky$^{a,b,}$\footnote{Alexander von
            Humboldt Fellow.}, D. M\"uller$^{a,c}$}

\vspace{18mm}

\centerline{\it ${^a}$Institut f\"ur Theoretische Physik, Universit\"at
                Regensburg}
\centerline{\it D-93040 Regensburg, Germany}
\centerline{\it ${^b}$Bogoliubov Laboratory of Theoretical Physics,
                Joint Institute for Nuclear Research}
\centerline{\it 141980, Dubna, Russia}
\centerline{\it ${^c}$Institute for Theoretical Physics, Center of
                Theoretical Science, Leipzig University}
\centerline{\it  04109 Leipzig, Germany}

\vspace{10mm}

\centerline{\bf Abstract}

\hspace{0.5cm}

Employing the operator algebra of the conformal group and the conformal
Ward identities, we derive the constraints for the anomalies of dilatation
and special conformal transformations of the local twist-2 operators in
Quantum Chromodynamics. We calculate these anomalies in the leading order
of perturbation theory in the minimal subtraction scheme. From the
conformal consistency relation we derive then the off-diagonal part of
the anomalous dimension matrix of the conformally covariant operators
in the two-loop approximation of the coupling constant in terms of
these quantities. We deduce corresponding off-diagonal parts of the
Efremov-Radyushkin-Brodsky-Lepage kernels responsible for the evolution of
the exclusive distribution amplitudes and non-forward parton distributions
in the next-to-leading order in the flavour singlet channel for the
chiral-even parity-odd and -even sectors as well as for the chiral-odd
one. We also give the analytical solution of the corresponding evolution
equations exploiting the conformal partial wave expansion.

\vspace{0.5cm}


\vspace{0.5cm}


\end{titlepage}

\section{Introduction.}

Quantum Chromodynamics (QCD) --- a relativistic local quantum field
theory based on the colour gauge group $SU_c(3)$ --- is well established
nowadays as a microscopic theory of strong interaction. However, due to
presently unsolved bound state problem, which remains a challenge for
future studies, it is not possible to predict measurable observables from
first principles. The standard approach which overcomes this difficulty
is based on factorization theorems \cite{CSS89} which allow to
separate soft and hard physics in a given process. The soft part is
described by expectation values of appropriate operators sandwiched
between hadronic states\footnote{Or more general correlation functions
of the elementary field operators in cases when the operator product
expansion does not exist.} supplied with appropriate evolution equations
which govern their scale dependence. Due to lack of any reliable
non-perturbative machinery, the former part can be fixed only from
experimental data. At the same time the hard subprocess and the
evolution of the soft function can be systematically calculated with
the help of perturbation theory. The success of perturbative QCD
in predicting the momentum dependence of hadronic observables
serves as the main and the most important argument for the correctness
of the theory. Therefore, one is interested in increasing the accuracy
of the theoretical predictions for their scale violation.

In general the calculations carried out beyond leading order
in the coupling constant are very difficult and require substantial
computer power. There are continuous advances in higher loop
calculations of inclusive processes. For instance, the
Dokshitzer-Gribov-Lipatov-Altarelli-Parisi (DGLAP) evolution
kernels \cite{GriLip72,AltPar77,Dok77} for the twist-2
parton distribution functions (see Refs.\ \cite{BFKL85,ArtMek90}
for transversity) are known in next-to-leading order (NLO)
\cite{Fur80,Nee96,Vog97} and the calculations for the
next-to-next-to-leading order (NNLO) are in progress. For exclusive
processes no comparable technical effort was invested so far.

For exclusive reactions at large momentum transfer the soft
physics is contained in distribution amplitudes (DA)
\cite{BroLep80,EfrRad80,DunMue80} or more general non-forward parton
distributions \cite{GriLevRys83,GeyDitHorMueRob88,Ji96,Rad96}, which
depend on the momentum fraction $x$ carried by the struck parton, and
possibly other kinematical variables such as the skewedness parameter
of the process. The scale dependence of the mesonic DA is governed by
the Efremov-Radyushkin-Brodsky-Lepage (ER-BL) evolution equation
\cite{EfrRad80,EfrRad78,BroLep79}
\begin{equation}
\label{ER-BLequation}
\mu^2\frac{d}{d\mu^2} \phi (x,\mu) =
\int_{0}^{1} dy V\left(x, y | \alpha_s(\mu) \right) \phi (y,\mu),
\end{equation}
where the evolution kernel $V (x, y | \alpha_s ) = \frac{\alpha_s}{2\pi}
V^{(0)} (x, y) + \left( \frac{\alpha_s}{2\pi} \right)^2 V^{(1)} (x, y)
+ \dots$ is given as a series in the coupling, and each expansion
coefficient is calculable in perturbation theory. Note that due to the
fact that the same light-ray operators enter the definition of DA's
and non-forward distributions, the corresponding evolution kernels are
in one-to-one correspondence \cite{GeyDitHorMueRob88}. In leading
order (LO) the calculations are quite straightforward since there exist
different reliable formalisms for the evaluation of non-forward
evolution kernels in the light-cone position as well as light-cone fraction
representations. On the other hand the direct calculation beyond
leading order turns out to be far from being a simple task which
has only been solved for the evolution kernel in the non-singlet
sector \cite{DitRad84,MikRad85,Sar84} and was checked numerically
in Ref.\ \cite{Kat85}. The most promising approach for two-loop
calculations developed in Ref.\ \cite{MikRad85} can be used for
future explicit NLO computation in the singlet channel. However, the
analytical solutions of the evolution equations for the DA
\cite{EfrRad78,BroFriLepSac80,MikRad86,Mue95} and non-forward
distributions \cite{Rad96,BelGeyMueSch97,ManPilWei97} are known in terms
of the anomalous dimensions $\gamma_{jk}$ of the, so-called, twist-2
conformal composite operators. They can be obtained from the evolution
kernels by forming moments with respect to Gegenbauer polynomials
\begin{equation}
\label{MomGegen}
\int_0^1 dx\, C^\nu_j (2x - 1) V( x , y | \alpha_s)
= - \frac{1}{2}
\sum_{k=0}^j \gamma_{jk}(\alpha_s) C^\mu_k (2y - 1),
\end{equation}
where the numerical value of the indices $\mu$ and $\nu$ depends on the
channel under consideration. Unfortunately, beyond LO this calculation
is very cumbersome and could not be performed analytically (only the first
few entries of this matrix have been evaluated numerically
\cite{MikRad86,KMR86}).

Owing to the difficulties we have sketched, every field-theoretical
idea is welcome which can help us to face the above mentioned problems
in a more economical way. On the first glance one could think that the
field-theoretical background of QCD has already been well studied in
the early days. However, after more than two decades of computation
experience it is desirable to have a new look on certain issues. In this
paper we apply an old idea of conformal symmetry\footnote{Here we warn
the reader from simple minded understanding of this terminology. From
the consequent discussion it will be clear that our point of view is more
sophisticated. It provides us with a powerful machinery, which enables us
to predict certain quantities in a very simple manner.}, which arises
from the generalization of the Poincar\'e group up to the fifteen
parameter conformal group isomorphic to $SO(4,2)$ --- the largest
possible group of transformations which leave light cone invariant.
Besides translations and four-dimensional rotations it contains
dilatations and special conformal transformations \cite{Kas62,MacSal69}.

In the massless case the classical QCD action is invariant under
infinitesimal conformal transformations. The restrictions coming from
conformal symmetry in massless scalar and gauge field theories were
studied already more than three decades ago when it was believed that this
symmetry is broken only softly \cite{Kas62,MacSal69,FerGriGat71a,Cre72}.
To give an example for the power of conformal symmetry, we remind
the reader that in a conformal invariant theory the functional form
of both elementary Green functions and current correlators is severely
constrained: two- and three-point functions are fixed completely up to
their normalization while $n$-point ($n \geq 4$) correlators are expressed
in terms of a single function of anharmonic ratios \cite{Pol71}.

Unfortunately, after quantization the theory is not conformal invariant even
in the massless limit. The symmetry is broken due to anomalies related to
the ultraviolet (UV) divergencies \cite{Jac70,Par72,Nie73,Sar74,Nie75}
which arise when one goes beyond the tree level. The regularization of
these divergencies introduces a dimensional parameter which breaks conformal
symmetry and this breaking still remains after renormalization and
removing of the regularization. It shows up in the trace anomaly of the
energy-momentum tensor \cite{AdlColDun77,ColDunJog77,Nie77,Min76}.
Getting rid of the unphysical contributions in the trace (see below
Eq.\ (\ref{TraceAnomaly})) such as BRST-exact operators, which originate from
the gauge fixing and ghost Lagrangians, and equations of motion we end up
finally with a symmetry breaking term proportional to the $\beta$-function.
If one assumes the existence of a non-perturbative fixed point $\g^\ast$
with $\beta (\g^\ast)=0$, the theory would be conformal invariant. Formally
we speak about the hypothetical conformal limit of the theory and set
$\beta$ equal to zero. Thus, the trace anomaly controls the conformal
symmetry breaking in the Ward identities of the elementary fields
operators \cite{Par72}. However, this is by no means true when one is
interested in more complicated Green functions, for instance, with
composite operator insertion, since this one-to-one correspondence between
both anomalies is lost. This means that in the conformal Ward identities
(CWIs) the breaking of the special conformal symmetry is not controlled
by the dilatation anomaly alone. Instead, new anomalies enter the game.

This should sound a warning from the naive use of the conformal symmetry
arguments. Indeed, in the past its application has led to a conflict between
its predictions and explicit calculations beyond leading order
\cite{BroDamFriLep86}. For instance, in the above mentioned example
of the renormalization of the conformal Ward identities with a composite
operator insertion there arises a new symmetry breaking term induced by the
UV divergencies occurring in the operator product of the composite
operator and the trace anomaly. Such a term is not proportional to the
$\beta$-function.

To study this breaking, we will construct Ward identities for the Green
functions of the elementary fields with conformal operator insertion
which will serve as a main tool for the derivation of constraints for
the anomalous dimensions beyond leading order. In general
the renormalization matrix of the twist-2 conformal operators is
triangular due to Lorentz invariance. In leading order it is
diagonal so that conformally covariant operators do not mix under
renormalization, i.e.\ $\gamma_{jk}^{(0)} =\delta_{jk}\gamma^{(0)}_j$
\cite{EfrRad80,BroFriLepSac80,Mak81,Ohr82,CraDobTod85}. Its eigenvalues
$\gamma_{j}$ coincide with the Mellin moments of the DGLAP kernels.
Therefore, the diagonal part\footnote{The formula given below is very
schematic and valid for direct ($QQ$ or $GG$) channels. For the
exhaustive treatment of this problem the reader is referred to the
paper \cite{BelMue98}.} of the exclusive evolution kernel can be
obtained by the transformation \cite{BelMue98}
\begin{equation}
V^{\rm D} ( x, y ) = \int_{0}^{1} dz P (z)
\sum_{j = 0}^{\infty} \frac{w (y | \nu)}{N_j(\nu)}
C^\nu_j (2x - 1) z^j C^\nu_j (2y - 1),
\end{equation}
where  $N_j(\nu)= 2^{ - 4 \nu + 1 }
\frac{ \Gamma^2 (\frac{1}{2}) \Gamma ( 2 \nu + j )}{\Gamma^2 (\nu)
( \nu + j ) j! }$ and $w (y | \nu) = (y \bar y)^{\nu-1/2}$ are the
normalization and weight factors, respectively. Thus, conformal
covariance at tree level is sufficient for the diagonality of
the one-loop anomalous dimension matrix. However, beyond
tree level conformal symmetry is broken even for $\beta = 0$. Of
course, the scaling dimensions of the operators are changed by
the anomalous ones but this shift does not ensure the covariance
of the renormalized operators under special conformal
transformation in a given renormalization scheme
\cite{Mue91a,Mue94,Mue97a,BelMue98}. This fact is easy to understand,
while the breaking of dilatation is determined in LO only by the
leading logarithmic term, the breaking of special conformal symmetry
is caused by both the logarithmic term and the remaining constant one.
Since the logarithmic independent term is fixed by the normalization
condition, which is implicitly given in the MS scheme, we can require
a normalization which ensures conformal covariance of the
renormalized operators\footnote{Exploiting this fact together with the
conformal operator product expansion we had, recently, the opportunity
to give several predictions for some non-forward processes
\protect\cite{Mue97a,BelMue97a,BelSch98}.} \cite{Mue97a,BelMue97a}.
This procedure is equivalent to a finite renormalization and provides
the, so-called, conformally covariant scheme. Therefore, the
renormalized conformally covariant operators at LO possess a
diagonal anomalous dimension matrix in two-loop approximation, up to a
term proportional to the $\beta$-function. Obviously, the non-diagonal
anomalous dimension matrix in the MS scheme in NLO can be obtained by
${\cal O}(\alpha_s)$ finite renormalization of the two-loop forward
anomalous dimensions. When translated into the language of evolution
kernels the whole result reads now:
\begin{equation}
V (x, y| \alpha_s)
= V^{\rm D} (x, y | \alpha_s)
+ V^{\rm ND} (x, y | \alpha_s), \quad
V^{\rm ND} (x, y | \alpha_s)
= \left(\frac{\alpha_s}{2\pi}\right)^2
V^{{\rm ND}(1)} (x , y) + {\cal O}(\alpha_s^3) .
\end{equation}
In other words the off-diagonal part $V^{{\rm ND}(1)}$ of the kernel is
a quasi-one-loop object and can be calculated in this order. This entry
will be evaluated in the present paper.

While in previous studies we restricted ourselves to the Abelian case, we
extend in this paper the conformal machinery to the flavour singlet sector
in QCD. Consequent presentation runs as follows. In the next section we
summarize some basic formulae. In section 3 we derive the renormalized CWI
analogous to the ones considered in Refs.\ \cite{Nie73,Sar74,Nie75} for the
composite operators in the dimensionally regularized theory. We do not rely
there on any assumption about symmetry breaking. The commutator constraints
for special conformal and anomalous dimension matrices are established
in section 4. Section 5 is devoted to calculation of the special conformal
anomaly kernels in both light-cone position and fraction representations.
Their transformation to the language of Gegenbauer moments is given in
section 6. The subsequent section contains explicit formulae for two-loop
anomalous dimensions as well as ER-BL kernels. Then we give the general
formalism for the solution of the evolution equation in the required
approximation. In the final section 8 we give our conclusions. To clarify
the presentations given in the body of the paper we add three technical
appendices. In the first one we list all Feynman rules required in the
calculation of the conformal anomalies. In appendix B we sketch the
light-cone position formalism for the computation of the leading order
evolution kernels. We mainly focus our attention on the most complicated
gluonic sector. The last appendix summarizes formulae for the
hypergeometric functions which appear in the evaluation of the Gegenbauer
moments of the LO evolution kernels.

\section{Preliminary remarks.}

To start with, let us recall some basic equations of QCD. The dimensionally
regularized and renormalized Lagrangian is given by
\begin{eqnarray}
\label{YM-renormalized}
{\cal L} &=&  Z_2 \bar\psi i \!\not\! \partial \psi
+ \bar Z_1 \mu^\epsilon \g \bar\psi\!\not\!\!B^a t^a \psi \nonumber\\
&-& \frac{Z_3}{4} \left(F^a_{\mu\nu}\right)^2
- \frac{1}{2} \mu^\epsilon \g Z_1 \ f^{abc} F^a_{\mu\nu}
B^b_\mu B^c_\nu - \frac{Z_4}{4} \mu^{2 \epsilon} \g^2
\left( f^{abc} B^b_\mu B^c_\nu \right)^2
- \frac{1}{2\xi} \left( \partial_\mu B^a_\mu \right)^2 \nonumber\\
&+& \widetilde Z_3 \ \partial_\mu \bar\omega^a \partial_\mu \omega^a
+ \mu^\epsilon \g \widetilde Z_1 \ f^{abc}
\partial_\mu \bar\omega^a B^b_\mu \omega^c,
\end{eqnarray}
where $d = 4 - 2 \epsilon$ is the space-time dimension and $\mu$ is a mass
parameter with dimension one introduced in order to keep the coupling
constant dimensionless. We introduce here the symbol $F^a_{\mu\nu}$
as a shorthand notation for the Abelian part of the full QCD field
strength tensor $G^a_{\mu\nu} = F^a_{\mu\nu} + \mu^\epsilon \g
X f^{abc} B^b_{\mu}B^c_{\nu}$. The canonical dimension of the elementary
fields are the following: $d^{\rm can}_\psi = \frac{3}{2} - \epsilon$
for fermions, $d^{\rm can}_G = 1 - \epsilon$ for gluons,
$d^{\rm can}_{\bar\omega} = d - 2$ and $d^{\rm can}_{\omega} = 0$
for the anti-ghost and ghost fields, respectively. This Lagrangian is
invariant under the following renormalized BRST-transformations:
\begin{eqnarray}
&&\delta^\BRST \psi =
- i \mu^\epsilon \g \widetilde{Z}_1 \omega^a t^a \psi \delta\lambda,
\quad
\delta^\BRST B^a_\mu =
\widetilde{Z}_3 D_\mu \omega^a \delta\lambda, \nonumber\\
&&\delta^\BRST \omega^a =
\frac{1}{2} \mu^\epsilon \g \widetilde{Z}_1 f^{abc}
\omega^b\omega^c \delta\lambda,
\quad
\delta^\BRST \bar{\omega}^a =
\frac{1}{\xi} \partial_\mu B_\mu^a \delta\lambda,
\end{eqnarray}
where $\delta\lambda$ is a renormalized Grassman variable.
The covariant derivative is defined as follows $D_\mu = \partial_\mu
-  i \mu^\epsilon \g X T^a B^a_\mu$, where $T^a$ is the generator in
the fundamental ($T^a \phi_i = t^a_{ij} \phi_j$) or adjoint ($T^b \phi^a
= i f^{abc} \phi^c$) representations depending on the object it
is acting on. The Ward-Takahashi identities imply the following
relations between the renormalization constants $Z_1 Z^{-\frac{3}{2}}_3
= Z^{\frac{1}{2}}_4 Z^{-1}_3 = \widetilde Z_1 \widetilde Z^{-1}_3
Z^{-\frac{1}{2}}_3 = \bar Z_1 Z_2^{-1} Z_3^{-\frac{1}{2}}$. In the
MS scheme $Z$-factors are defined as Laurent series in $\epsilon$:
$Z(\g, \epsilon)=1 + Z^{[1]}(\g, \epsilon)/\epsilon +
Z^{[2]}(\g, \epsilon)/\epsilon^2 + \dots$. For our consequent
discussion we take the renormalization constants $Z_2$, $Z_3$,
$\widetilde Z_3$ and $X \equiv Z_1 Z^{-1}_3 = \widetilde Z_1
\widetilde Z^{-1}_3 = \bar Z_1 Z_2^{-1}$ as independent ones.

The renormalization group coefficients and the anomalous dimensions of the
fields are
\begin{eqnarray}
\beta_\epsilon (\g )
= \mu \frac{d \g }{d \mu}
= - \epsilon \, \g + \beta (\g ),
\quad
\sigma = \mu \frac{d}{d \mu} \ln \xi = -2 \gamma_G,
\quad
\gamma_\phi = \frac{1}{2} \mu \frac{d}{d \mu} \ln Z_\phi,
\end{eqnarray}
($\phi$ runs over parton species entering the Lagrangian $\{ \phi =
\psi, G, \bar\omega \}$) where $Z_\psi = Z_2,\ Z_G = Z_3,\ Z_{\bar\omega}
= \widetilde Z_3$. These coefficients are determined by the residues of
the corresponding $Z$-factors \cite{Hoo73,Col74} and in lowest order
approximation, $\frac{\beta}{\g} = \frac{\alpha_s}{4\pi} \beta_0
+ \left( \frac{\alpha_s}{4\pi} \right)^2 \beta_1 + \dots$, $\gamma_\phi
= \frac{\alpha_s}{2\pi} \gamma^{(0)}_\phi + \dots$, they are given by
\begin{equation}
\beta_0 = \frac{4}{3} T_F N_f - \frac{11}{3}C_A ,
\quad
\gamma^{(0)}_G (\xi) = \frac{2}{3} T_F N_f
+ \frac{C_A}{4} \left( \xi - \frac{13}{3} \right),
\quad
\gamma^{(0)}_\psi (\xi) = \frac{\xi}{2} C_F,
\end{equation}
where the group theoretical factors are $C_A=3$, $C_F=4/3$, and $T_F=1/2$
for $SU_c(3)$.

In the following we are interested in the renormalization of the
conformal operators in the flavour singlet sector, which we define as
two-dimensional vector:
\begin{equation}
{\cal O}_{jl}
= \left( { {^Q\!{\cal O}_{jl}} \atop {^G\!{\cal O}_{jl}} } \right) .
\end{equation}
The group-theoretical construction of these local operators was given
sometimes ago in Refs.\ \cite{Mak81,Ohr82} (see also Ref.\ \cite{BalBra89}
for a non-local version of conformal operators)
\begin{equation}
\label{treeCO}
\left\{\!\!\!
\begin{array}{c}
{^Q\!{\cal O}^V} \\
{^Q\!{\cal O}^A}
\end{array}
\!\!\!\right\}_{jl}
\!=
\bar{\psi} (i \partial_+)^l\!
\left\{\!\!\!
\begin{array}{c}
\gamma_+ \\
\gamma_+ \gamma_5
\end{array}
\!\!\!\right\}
\!C^{\frac{3}{2}}_j\!
\left( \frac{\stackrel{\leftrightarrow}{D}_+}{\partial_+} \right)
\!\psi, \
\left\{\!\!\!
\begin{array}{c}
{^G\!{\cal O}^V} \\
{^G\!{\cal O}^A}
\end{array}
\!\!\!\right\}_{jl}
\!=
G_{+ \mu} (i \partial_+)^{l-1}\!
\left\{\!\!\!
\begin{array}{c}
g_{\mu\nu} \\
i\epsilon_{\mu\nu-+}
\end{array}
\!\!\!\right\}
\!C^{\frac{5}{2}}_{j - 1}\!
\left(
\frac{\stackrel{\leftrightarrow}{D}_+}{\partial_+}
\right)
\!G_{\nu +},
\end{equation}
where $\partial \!= \stackrel{\rightarrow}{\partial}
\!\!+\!\! \stackrel{\leftarrow}{\partial}$
and  $\stackrel{\leftrightarrow}{D}
= \stackrel{\rightarrow}{D} - \stackrel{\leftarrow}{D}$. The $+$ and $-$
components are obtained by contraction with the two light-like vectors
$n$ and $n^*$, such that $n^2 = n^{*2} = 0$ and $nn^* = 1$. The indices
of the Gegenbauer polynomials are determined by the canonical dimensions
$d^{\rm can}_\phi$ and the spins $s_\phi$ of the fields: $\nu_\phi =
d^{\rm can}_\phi + s_\phi
- 1/2$.

The composite operators (\ref{treeCO}) form an infinite representation of
the collinear conformal algebra at the tree level. This subalgebra $SU(1,1)$
of the whole conformal algebra contains the generators
\begin{eqnarray}
{\cal P}_+ = n_\mu {\cal P}_\mu , \
{\cal M}_{-+}=n^*_\mu {\cal M}_{\mu\nu} n_\nu , \
{\cal D} , \
{\cal K}_- = n^*_\mu {\cal K}_\mu .
\end{eqnarray}
The total momentum  ${\cal P}_+$ is a raising, the special conformal
generator ${\cal K}_-$ is a lowering and the dilatation ${\cal D}$ and
the Lorentz ${\cal M}_{-+}$ generators are diagonal operators:
\begin{eqnarray}
\label{confVaria}
\delta^P_+ \, {\cal O}_{jl}
&=& i [ {\cal O}_{jl}, {\cal P}_+ ]_-
= i\, {\cal O}_{jl+1} , \hspace{0.8cm}
\delta^M_{-+} \, {\cal O}_{jl}
= i [ {\cal O}_{jl}, {\cal M}_{-+} ]_-
= -( l + 1 ) {\cal O}_{jl} , \\
\label{confVaria-1}
\delta^S \, {\cal O}_{jl}
&=& i [{\cal O}_{jl}, \D ]_-
= - ( l + 3 ) {\cal O}_{jl}, \quad
\delta^C_- \, {\cal O}_{jl}
= i [ {\cal O}_{jl}, \K_- ]_-
= i\, a (j,l) {\cal O}_{jl-1} ,
\end{eqnarray}
with $a_{jk}(l) = \delta_{jk} \cdot a(j,l)$ and $a (j, l) = 2
(j - l)( j + l + 3)$. There is one Casimir operator of the collinear
conformal algebra, which is given by ${J\!\!\!\!J}^2 \, =
\frac{1}{2} {\cal P}_+ {\cal K}_- - \frac{1}{4} ( {\cal D}
+ {\cal M}_{-+} )^2 - \frac{i}{2} ( {\cal D} + {\cal M}_{-+} )$. The
operators introduced above possess conformal spin $j + 1$:
${J\!\!\!\!J}^2 \, {\cal O}_{jl} = (j + 1)(j + 2) {\cal O}_{jl}$, spin
$l + 1$ and scale dimension $l + 3$.

To study the conformal anomalies we consider the CWI for the connected
Green functions\footnote{It is straightforward to perform Legendre
transformation to derive the CWI for the 1-particle irreducible Green
functions. We will avoid this technical step, since we can obtain all
desired results from the CWI of the connected Green functions.} defined
by the formula
\begin{equation}
\label{Low-formula}
\langle [{\cal O}_{jl}] \X \rangle =
\frac{\int D\phi [{\cal O}_{jl}] \phi(x_1)\dots \phi(x_k) e^{i S}}
{\int D\phi e^{iS}},
\end{equation}
$\X$ is a product of elementary fields $\X = \prod_{i=1}^N \phi (x_i)$
at different space-time points. The integration is weighted with the
factor $e^{iS}$, where $S$ is the QCD action determined in terms of
the Lagrangian (\ref{YM-renormalized}) $S = \int d^d x {\cal L}$, and
\begin{equation}
\label{renCO-OP}
[ {\cal O}_{jl} ]
= \sum_{k = 0}^{j} Z_{jk}(\epsilon, \g) {\cal O}_{kl}
\end{equation}
are the renormalized composite operators. Here we neglect possible
gauge-variant counterterms, since they give no contribution to the
physical sector of the theory. The four entries of the matrix
${\!^{AB}Z}_{jk}(\epsilon, \g)$ with $A,B=\{Q,G\}$ are triangular
and independent of $l$ due to Poincar\'e invariance. In the MS
scheme the expansion ${\!^{AB}Z}_{jk}(\epsilon, \g) = \delta^{AB}
\delta_{jk} + {\!^{AB}Z}^{[1]}_{jk}(\g)/\epsilon + \cdots$ is valid
\cite{Hoo73,Col74}. Since the regularization is manifest in the action,
the true conformal Ward identities, i.e.\ those which are valid in the
interacting theory, can be obtained by a formal conformal variation of
Eq.\ (\ref{Low-formula}):
\begin{equation}
\label{CWI}
\langle [{\cal O}_{jl}] \delta \X \rangle
=-\langle \delta [{\cal O}_{jl}] \X \rangle -
\langle i [{\cal O}_{jl}] \delta S \X \rangle .
\end{equation}
Here the LHS contains a differential operator acting on a renormalized
Green function and, therefore, it is finite by itself. Thus, the same
property should be satisfied by the RHS. Because of Poincar\'e
invariance, the variations $\delta_+^P S$ and $\delta_{-+}^M S$
vanish identically and thus, with Eq.\ (\ref{confVaria}) the
corresponding Ward identities read:
\begin{equation}
\label{WI-Poincare}
\langle [{\cal O}_{jl}] \delta_+^P \X \rangle
=-i\langle  [{\cal O}_{jl+1}] \X \rangle \qquad
\langle [{\cal O}_{jl}] \delta_{-+}^M \X \rangle
=(l+1) \langle  [{\cal O}_{jl}] \X \rangle.
\end{equation}
In the case of conformal transformations, i.e.\ $\delta = \{ \delta^S,
\delta^C_- \}$, the variations of the action do not vanish anymore. The
operator product $[{\cal O}_{jl}] \delta S$ appearing on the RHS of
Eq.\ (\ref{CWI}) is infinite. The removal of the UV divergencies induces
the conformal anomalies, i.e.\ anomalous dimension and special conformal
anomaly matrices.

Before we start our study of the conformal variation of the action, we
write down the variation of the composite operators when the interaction is
switched on. Even in the interacting theory it is convenient to deal with
scale dimensions which are equal to the canonical ones in 4-dimensional
space-time, i.e.\ $d_G=1$, $d_\psi = \frac{3}{2}$. Therefore, the
infinitesimal conformal transformations of the renormalized
operators (\ref{renCO-OP}) are simply given by
\begin{equation}
\label{VarOper}
\delta^S \, [{\cal O}_{jl}]
= - ( l + 3 ) [{\cal O}_{jl}] , \qquad
\delta^C_- \, [{\cal O}_{jl}] =
i\, \sum_{k = 0}^{j}
\left\{ \hat Z \, \hat a (l) \hat Z^{-1} \right\}_{jk}
[{\cal O}_{k l-1}].
\end{equation}

The choice $d_G=1$, $d_\psi = \frac{3}{2}$, $d_{\bar{\omega}} = d - 2$ and
$d_\omega = 0$ ensures that the conformal variation of the action can be
expressed by well classified operators: gauge invariant operators, BRST
exact operators, and equation of motion (EOM) operators. A further
advantage of this set of scale dimensions is that it allows to use Jackiw's
conformal covariant transformation law for the gauge fields \cite{Jac78}.
After some algebra, the final result can be written as
\begin{eqnarray}
\label{VarAction}
\delta^S S = \int d^d x\ \Delta (x) , \quad
\delta^C_\nu S = \int d^d x\ 2 x_\nu\, \Delta (x) ,
\end{eqnarray}
where the integrand $\Delta (x)$ can be expressed in terms of the
regularized but unrenormalized trace anomaly of the energy-momentum
tensor ${\mit\Theta}_{\mu\nu}$:
\begin{eqnarray}
\label{TraceAnomaly}
\Delta (x) &=& - {\mit\Theta}_{\mu\mu} (x)
- d_G {\mit \Omega}_{G} (x)
- d_\psi {\mit \Omega}_{\bar\psi\psi} (x)
- d_{\bar\omega} {\mit \Omega}_{\bar\omega} (x) \nonumber\\
&=& \epsilon\,
\left\{
{\cal O}_A (x) + {\cal O}_B (x)
+ {\mit \Omega}_{\bar \omega} (x)
- {\mit \Omega}_{\bar\psi\psi} (x)
\right\} + (d - 2)\,\partial_\mu {\cal O}_{B\mu}(x) .
\end{eqnarray}
Here we introduce the following set of operators
\begin{eqnarray}
{\cal O}_A (x) = \frac{Z_3}{2} \left( G^a_{\mu\nu} \right)^2, \ \
{\cal O}_B (x) = \frac{\delta^{\rm BRST}}{\delta\lambda}
\bar{\omega}^a\partial_\mu B_\mu^a, \ \
{\cal O}_{B\mu} (x) = \frac{\delta^{\rm BRST}}{\delta\lambda}
\bar{\omega}^a B^a_\mu,
\end{eqnarray}
as well as EOM operators
\begin{equation}
{\mit \Omega}_G (x) = B^a_\mu \frac{\delta S}{\delta B^a_\mu},
\quad
{\mit \Omega}_{\bar\psi\psi} (x)
= \frac{\delta S}{\delta \psi} \psi
+ \bar\psi \frac{\delta S}{\delta \bar\psi},
\quad
{\mit \Omega}_{\bar \omega} (x)
= \bar \omega^a \frac{\delta S}{\delta \bar \omega^a}.
\end{equation}

It is desirable to express these variations of the action in terms of
renormalized operators as this allows to neglect then terms proportional
to $\epsilon $. For this reason the renormalization problem of the above
mentioned operators have to be solved \cite{JogLee75}. The only gauge
invariant operator ${\cal O}_A$ (class A) is of twist-4 and needs
counterterms of class B operators, given as BRST-exact ones, and equation
of motion operators (class C). The twist-4 class B operators are
${\cal O}_B$ and the total derivative of the twist two operator
$\partial_\mu {\cal O}_{B\mu}$. Since there is no other class B operator
which they can mix with, only class C operators are needed as counterterms
for the class B operators. This set of twist-4 and twist-2 operators
is closed under renormalization. First, we discuss the renormalization of
the twist-2 operator ${\cal O}_{B\mu} $. Since there is no further class B
or class C operator with dimension three and the same quantum numbers we
may conclude that this operator is renormalized ${\cal O}^\nu_B (x)
= [ {\cal O}^\nu_B (x) ]$. It has vanishing physical matrix elements
at non-zero momentum transfer \cite{Col84} while at zero momentum
transfer it vanishes identically as it contains no poles.
The renormalization problem of the twist-4 operators at zero
momentum can be solved by a direct calculation of the following
differential vertex operator insertions\footnote{Throughout the
paper, to simplify notations, we use the following conventions:
for unintegrated operator insertions and EOM operators we keep the
dependence on the space-time point explicit ${\cal O} (x) $; for the
integrated quantities with weight function 1 we just omit this
dependence ${\cal O} \equiv \int d^d x {\cal O} (x)$, while for weight
$2x_-$ we use ${\cal O}^- \equiv \int d^d x \, 2 x_- {\cal O} (x)$.}:
\begin{equation}
\label{def-DIVOI}
[\Delta^\xi] \equiv \xi \frac{\partial}{\partial\xi} S
= \frac{1}{2} \Bigl\{ \left[{\cal O}_B\right]
+ {\mit \Omega}_{\bar \omega} \Bigr\}, \quad
[\Delta^\g] \equiv \g \frac{\partial}{\partial\g} S
= \left[{\cal O}_A\right]
+ \left[{\cal O}_B\right]
+ {\mit \Omega}_G
+ {\mit \Omega}_{\bar \omega} .
\end{equation}
We tentatively expressed in these equations the derivatives of
the renormalization constants in terms of the derivatives of
independent set. From the finiteness of the differential vertex
operator insertions the counterterms of $[ {\cal O}_A (x) ]$ and
$[ {\cal O}_B (x) ]$ have been fixed up to total derivatives (here
$Z_\g = X Z_3^{-\frac{1}{2}}$):
\begin{eqnarray}
\left[ {\cal O}_A (x) \right]
&=&
\left( 1 + \g \frac{\partial\ln Z_\g }{\partial \g} \right)
{\cal O}_A (x)
+ \left( \g \frac{\partial\ln X}{\partial \g}
- 2 \xi \frac{\partial\ln X}{\partial \xi} \right)
\Bigl\{ {\cal O}_B (x)
+ {\mit \Omega}_{\bar \omega} (x)
+ {\mit \Omega}_G (x) \Bigr\} \\
&&+ \left( \g \frac{\partial\ln \widetilde Z_3}{\partial \g}
- 2\xi\frac{\partial\ln\widetilde Z_3 }{\partial \xi}\right)
{\mit \Omega}_{\bar \omega} (x)
+ \frac{1}{2}
\left( \g \frac{\partial\ln \widetilde Z_2}{\partial \g}
- 2\xi\frac{\partial\ln\widetilde Z_2 }{\partial \xi}\right)
{\mit \Omega}_{\bar\psi\psi} (x)
+ \partial_\mu {\cal R}^\mu_A (x), \nonumber\\
\left[ {\cal O}_B (x) \right]
&=&
\left( 1 + 2 \xi\frac{\partial\ln X}{\partial \xi} \right) {\cal O}_B (x)
+ 2 \xi\frac{\partial\ln X }{\partial\xi}
\Bigl\{
{\mit \Omega}_G (x)
+ {\mit \Omega}_{\bar \omega} (x)
\Bigr\}
+ \xi\frac{\partial\ln\widetilde Z_2 }{\partial \xi}
{\mit \Omega}_{\bar\psi\psi} (x) \\
&&+ 2 \xi\frac{\partial\ln\widetilde Z_3 }{\partial \xi}\
{\mit \Omega}_{\bar \omega} (x)
+ \partial_\mu {\cal R}^\mu_B (x) .\nonumber
\end{eqnarray}
Making use of the representation
\begin{eqnarray}
\epsilon = -\frac{\beta_\epsilon (\g)}{\g}
\left( 1 + \g\frac{\partial\ln Z_\g}{\partial \g} \right)
\end{eqnarray}
we can express
$\epsilon\, \{ {\cal O}_A (x) + {\cal O}_B (x) \}$ in terms of the
renormalized operators only
\begin{eqnarray}
\epsilon
\{ {\cal O}_A (x) + {\cal O}_B (x) \}
=
- \frac{\beta_\epsilon}{\g} [ {\cal O}_A (x) ]
\!\!&-&\!\! \left(
\frac{\beta_\epsilon}{\g} - \gamma_G
\right)
[ {\cal O}_B (x) ]
+ \left(
\gamma_G - \frac{\beta}{\g}
\right)
\Bigl\{
{\mit \Omega}_G (x) + {\mit \Omega}_{\bar \omega} (x)
\Bigr\} \nonumber\\
&+&\!\! \gamma_\psi {\mit \Omega}_{\bar\psi\psi} (x)
+ 2 \gamma_\omega {\mit \Omega}_{\bar\omega} (x).
\end{eqnarray}
Thus, inserting our findings into Eq.\ (\ref{TraceAnomaly}) we obtain the
renormalized anomaly:
\begin{eqnarray}
\label{trace-anomaly}
\Delta (x)
&=& - \frac{\beta_\epsilon}{\g} [ {\cal O}_A (x) ]
- \left(
\frac{\beta_\epsilon}{\g} - \gamma_G
\right)
\Bigl\{
[ {\cal O}_B (x) ]
+
{\mit \Omega}_{\bar\omega} (x)
\Bigr\}
+ \left( \gamma_G - \frac{\beta}{\g} \right)
{\mit \Omega}_G (x)  \nonumber\\
&&+ ( \gamma_\psi - \epsilon ) {\mit \Omega}_{\bar\psi\psi} (x)
+ 2 \gamma_\omega {\mit \Omega}_{\bar \omega} (x)
+ (d - 2) \partial_\mu [ {\cal O}_{B\mu} (x) ] .
\end{eqnarray}

\section{Renormalization of the Ward identities.}

Combining the results we have obtained so far in Eqs.\ (\ref{CWI}),
(\ref{VarOper}), (\ref{VarAction}) and (\ref{trace-anomaly}), we will
come to the desired conformal Ward identities. However, since the
products of composite operators possess UV divergencies, we have to
take care of their subtractive renormalization. The main problem
here is that no theorems for the renormalization of (gauge invariant)
operator products exist \cite{Jog77a}.

This section is devoted to the solution of this problem for the operator
products appearing in the CWI (\ref{CWI}). It is worth to mention that
in our previous work on Abelian gauge theory \cite{BelMue98} we discriminate
a rigorous treatment in favour of a symmetrical handling of the quark and
gluon sectors. Due to the non-Abelian character of the present consideration
it is helpful to deal with the well classified operators ${\cal O}_i$. Of
course, both approaches --- the one which we have used there and the other
we describe below --- lead to the same results for a theory with $U(1)$
symmetry group.

Fortunately, due to the fact that the products involved are composed of
gauge invariant conformal operators and differential vertex operator
insertions the form of the gauge-variant part can be fixed from the study
of the Green function (\ref{Low-formula}) with the help of the action
principle. So that finally we have for the unintegrated operators:
\begin{eqnarray}
\label{ren-A}
i[{\cal O}_A (x)] [{\cal O}_{j l}]
= i[{\cal O}_A (x){\cal O}_{j l}]
\!\!\!&-&\!\!\!
\delta^{(d)} (x) \sum_{k = 0}^{j}
\left\{ \hat Z_A \right\}_{jk}
[{\cal O}_{k l}]
-
\frac{i}{2} \partial_+ \delta^{(d)} (x) \sum_{k = 0}^{j}
\left\{ \hat Z_A^- \right\}_{jk}
[{\cal O}_{k l - 1}]
- \dots
\nonumber\\
\!\!\!&-&\!\!\!
\left(
\g \frac{\partial\ln X}{\partial\g}
- 2 \xi \frac{\partial\ln X}{\partial\xi}
\right)
B_\mu^a (x) \frac{\delta}{\delta B_\mu^a (x)}
[{\cal O}_{j l}] ,
\\
\label{ren-B}
i[{\cal O}_{B^\ast} (x)] [{\cal O}_{j l}]
= i[{\cal O}_{B^\ast} (x) {\cal O}_{j l}]
\!\!\!&-&\!\!\!
\delta^{(d)} (x) \sum_{k = 0}^{j}
\left\{ \hat Z_B \right\}_{jk}
[{\cal O}_{k l}]
- \frac{i}{2} \partial_+ \delta^{(d)} (x) \sum_{k = 0}^{j}
\left\{ \hat Z_B^- \right\}_{jk}
[{\cal O}_{k l - 1}]
- \dots \nonumber\\
\!\!\!&-&\!\!\!
\left(
2 \xi \frac{\partial}{\partial\xi} \ln X
\right)
B_\mu^a (x) \frac{\delta}{\delta B_\mu^a (x)}
[{\cal O}_{j l}] .
\end{eqnarray}
The ellipses stand for the higher derivatives counterterms which are
not of relevance for our present study. For convenience we introduced
here the operator combination $[{\cal O}_{B^\ast} (x) ] = [{\cal O}_B (x) ]
+ {\mit \Omega}_{\bar \omega} (x)$, which on tree level is the
gauge fixing term and thus require only counterterms proportional to
$\xi$. In the Landau gauge the operator product $[{\cal O}_{B^\ast} (x)]
[{\cal O}_{j l}]$ is finite, since all counterterms will vanish. Obviously,
the term containing the variation of the composite operators with respect
to the gauge field is gauge-variant (it cannot be decomposed in terms of
A, B or EOM operators). In the following subsection we show that the
appearance of this term ensures the correct renormalization group equation
for the composite operators. We also will fix the $Z$-matrices $\hat Z_A$
and $\hat Z_B$ in terms of known renormalization group coefficients.

Fortunately, by employing BRST invariance and naive power counting
we are able to show in general that the operator product
$[ {\cal O}_{jl}][{\cal O}_{B\mu} (x)]$ is renormalized:
\begin{eqnarray}
\label{ren-ext}
[ {\cal O}_{jl}][{\cal O}_{B\mu} (x)]
=
[ {\cal O}_{jl} {\cal O}_{B\mu} (x)].
\end{eqnarray}

Finally, we have also to deal with the product of composite operators
and EOM operators. For these products we do not use the MS prescription,
rather we perform a mere integration in the path integral which results
in equation:
\begin{eqnarray}
\label{ren-EOM}
i  [ {\cal O}_{jl}] {\mit \Omega}_\phi(x) =  i  [ {\cal O}_{jl}
{\mit \Omega}_\phi(x)]
-\phi(x) \frac{\delta}{\delta \phi(x)} [ {\cal O}_{jl}].
\end{eqnarray}
where it is obvious that
\begin{eqnarray}
\label{ren-EOM-1}
\langle [ {\cal O}_{jl} {\mit\Omega}_\phi(x)] \X \rangle
= i \langle
[ {\cal O}_{jl}] \phi (x) \frac{\delta}{\delta \phi(x)} \X
\rangle.
\end{eqnarray}
For instance, for the counterterms of the products containing the EOM
operator of the quark fields we have
\begin{eqnarray}
\int d^d x \left(
\bar\psi \frac{\delta}{\delta \bar\psi}
+
\frac{\delta}{\delta \psi} \psi
\right)
[{\cal O}_{jl}]
&=& 2 \sum_{k = 0}^{j}
\left\{
\hat Z \hat P_Q \hat Z^{-1}
\right\}_{jk} [{\cal O}_{kl}],
\\
\int d^d x 2x_-\left(
\bar\psi \frac{\delta}{\delta \bar\psi}
+
\frac{\delta}{\delta \psi} \psi
\right)
[{\cal O}_{jl}]
&=& 2 i \sum_{k = 0}^{j}
\left\{
\hat Z \hat b \hat P_Q \hat Z^{-1}
\right\}_{jk} [{\cal O}_{kl-1}].
\end{eqnarray}
We have used above the projectors on the quark and gluon sectors
\begin{eqnarray}
\hat P_Q =
\left( { 1 \ 0 \atop 0 \ 0 } \right),
\qquad
\hat P_G =
\left( { 0 \ 0 \atop 0 \ 1 } \right),
\end{eqnarray}
and the $\hat b$-matrix with elements given by
\begin{eqnarray}
\label{b-matrix}
b_{jk} (l, \nu)
&=& 2
\int_{0}^{1} d x
\frac{ w(x | \nu) }{ N_{k}( \nu ) }C_{k}^\nu ( 2x - 1 )
\left[ l - \frac{1}{2} ( 2x - 1 ) \frac{\partial}{\partial x} \right]
C_{j}^\nu ( 2x - 1 )
\nonumber\\
&=& 2\, \theta_{jk}
\left\{ ( l + k + 2 \nu ) \delta_{jk}
- [ 1 + (- 1)^{ j - k} ] ( k + \nu )
\right\} .
\end{eqnarray}
Due to the definition of the conformal operators (\ref{treeCO}), the
same matrix $b_{jk}(l)\equiv b_{jk}(l,3/2) =b_{j-1k-1}(l-1,5/2)$
appears in the quark as well as in the gluon channels.

\subsection{Scale Ward identity.}

The renormalization of the dilatation Ward identity is quite
straightforward in terms of the differential vertex operator insertions
(\ref{def-DIVOI}). Nevertheless, we present here an alternative
way, which, however, can also be used for the renormalization of
the special conformal Ward identities. Afterwards we show that the
appearing dilatation anomaly is just the anomalous dimension matrix
of the composite operators.

Taking into account the subtractive renormalization procedures in Eqs.\
(\ref{ren-A}), (\ref{ren-B}) and (\ref{ren-EOM}), we find with
Eqs.\ (\ref{TraceAnomaly}) and (\ref{trace-anomaly}) the following
equation for the operator product appearing in the Ward identity
(\ref{CWI}) (with $\delta\equiv\delta^S$) expressed via renormalized
insertions:
\begin{eqnarray}
i [ {\cal O}_{jl}] \delta^S S &=&
\sum_{k = 0}^{j}
\left\{ \frac{\beta_\epsilon}{\g} \hat Z_A+
\left( \frac{\beta_\epsilon}{\g} - \gamma_G \right) \hat Z_B
- 2 ( \gamma_\psi - \epsilon ) \hat Z \hat P_Q \hat Z^{-1}
\right\}_{jk} [ {\cal O}_{kl} ]
\nonumber\\
&& - \frac{\beta_\epsilon}{\g} i [ {\cal O}_{jl}{\cal O}_A] -
\left( \frac{\beta_\epsilon}{\g} - \gamma_G \right)
i [ {\cal O}_{jl} {\cal O}_{B^\ast}]
+ \left( \gamma_G - \frac{\beta}{\g} \right )
i [{\cal O}_{jl}{\mit \Omega}_{G}]
\\
&&+
( \gamma_\psi - \epsilon ) i [{\cal O}_{jl}{\mit \Omega}_{\bar\psi\psi}]
+ 2 \gamma_\omega
i [{\cal O}_{jl}{\mit \Omega}_{\bar{\omega}}] .
\nonumber
\end{eqnarray}
Inserting this result in the Ward identity (\ref{CWI}) we get
with Eq.\ (\ref{confVaria-1}) in the limit $\epsilon \to 0$:
\begin{eqnarray}
\label{DWI}
\langle [ {\cal O}_{jl} ] \delta^S \X \rangle
&=& \sum_{k = 0}^{j}
\left\{( l + 3 ) \hat 1+\hat\gamma \right\}_{jk}
\langle  [ {\cal O}_{kl} ] \X \rangle +  \frac{\beta}{\g}
\langle i [ {\cal O}_{jl} {\cal O}_A] \X \rangle
+ \left(\frac{\beta}{\g}-\gamma_G\right)
\langle i [ {\cal O}_{jl} {\cal O}_{B^\ast}] \X \rangle
\nonumber\\
&&
-\gamma_\psi
\langle i[ {\cal O}_{jl} {\mit \Omega}_{\bar\psi\psi} ] \X \rangle
-\left( \gamma_G - \frac{\beta}{\g} \right )
\langle i[ {\cal O}_{jl} {\mit \Omega}_{G} ] \X \rangle
- 2 \gamma_\omega
\langle i[ {\cal O}_{jl} {\mit \Omega}_{\bar\omega} ] \X \rangle,
\end{eqnarray}
where the finiteness of the LHS of the Ward identity ensures the
existence of the matrix
\begin{equation}
\label{def-delaAno}
\hat\gamma = \lim_{\epsilon\to 0}
\left\{ - \frac{\beta_\epsilon}{\g} \hat Z_A
- \left(\frac{\beta_\epsilon}{\g}-\gamma_G\right) \hat Z_B
+ 2 ( \gamma_\psi - \epsilon )
\hat Z \hat P_Q \hat Z^{-1}\right\} .
\end{equation}

It remains to show that the dilatation Ward identity (\ref{DWI}) is
nothing else than the  familiar Callan-Symanzik equation
\cite{Cal70,Sym70} and that $\hat\gamma$ is the anomalous dimension
matrix defined in the MS scheme. By differentiation of the renormalized
Green function with respect to the parameters $\g$ and $\xi$ we define
the differential vertex operator insertions $[{\cal O}_{jl}\Delta^i]$,
so that
\begin{eqnarray}
\label{DVOI-g}
i [ {\cal O}_{jl} ] [\Delta^\g ]
=
i [ {\cal O}_{jl} \, \Delta^\g ]
&-&
\sum_{k = 0}^{j}
\biggl\{
\g \frac{\partial \hat Z}{\partial \g} \hat Z^{-1} \nonumber\\
&+&
\left(
1 + \g \frac{\partial \ln X}{\partial \g}
\right)
\hat Z
\left( \hat 1
\int d^d x B_\mu^a (x) \frac{\delta}{\delta B_\mu^a (x)}
-
2 \hat P_G
\right) \hat Z^{-1}
\biggr\}_{jk}\!
[ {\cal O}_{kl} ] ,
\\
\label{DVOI-xi}
i [ {\cal O}_{jl} ] [\Delta^\xi ]
=
i [ {\cal O}_{jl} \, \Delta^\xi ]
&-&
\sum_{k = 0}^{j}
\biggl\{
\xi \frac{\partial \hat Z}{\partial \xi} \hat Z^{-1} \nonumber\\
&+&
\xi \frac{\partial \ln X}{\partial \xi}
\hat Z
\left( \hat 1
\int d^d x B_\mu^a (x) \frac{\delta}{\delta B_\mu^a (x)}
-
2 \hat P_G
\right) \hat Z^{-1}
\biggr\}_{jk}\!
[ {\cal O}_{kl} ] .
\end{eqnarray}

We learn from Eqs.\
(\ref{def-DIVOI},\ref{ren-A},\ref{ren-B},\ref{DVOI-g},\ref{DVOI-xi})
that the renormalization constants $\hat Z_i$ can be expressed in terms
of the renormalization matrix of conformal operators and read
\begin{eqnarray}
\hat Z_A &=&
\left(
\g \frac{\partial\hat Z}{\partial\g}
- 2 \xi \frac{\partial\hat Z}{\partial\xi}
\right) \hat Z^{-1} - 2 \hat Z \hat P_G \hat Z^{-1}
- 2 \left(
\g \frac{\partial\ln X}{\partial\g}
- 2 \xi \frac{\partial\ln X}{\partial\xi}
\right) \hat Z \hat P_G \hat Z^{-1} , \\
\hat Z_B &=&
2 \xi \frac{\partial\hat Z}{\partial\xi} \hat Z^{-1}
- 4 \left(
\xi \frac{\partial\ln X}{\partial\xi}
\right) \hat Z \hat P_G \hat Z^{-1} .
\end{eqnarray}

Inserting these findings into the renormalized dilatation Ward identity
(\ref{DWI}) provides us the well-known form of this Ward identity:
\begin{eqnarray}
\label{DWI-1}
\langle [ {\cal O}_{jl} ] \delta^S \X \rangle
&=& \sum_{k = 0}^{j}
\left\{
( l + 3 ) \hat 1 + \hat\gamma \right\}_{jk}
\langle [ {\cal O}_{kl} ] \X \rangle
+ \frac{\beta}{\g}
\langle i [ {\cal O}_{jl} \Delta^\g ] \X \rangle
+  \sigma
\langle i[ {\cal O}_{jl} \Delta^\xi ] \X \rangle
\nonumber\\
&&- \gamma_\psi
\langle i[ {\cal O}_{jl} {\mit \Omega}_{\bar\psi\psi} ] \X \rangle
- \gamma_G
\langle i[ {\cal O}_{jl} {\mit \Omega}_G ] \X \rangle
-2 \gamma_\omega
\langle i[ {\cal O}_{jl} {\mit \Omega}_{\bar \omega} ] \X \rangle,
\end{eqnarray}
and from Eq.\ (\ref{def-delaAno}) follows the familiar definition for
the anomalous dimension matrix:
\begin{equation}
\hat\gamma \equiv \hat\gamma_Z + 2\gamma_G \hat P_G + 2 \gamma_\psi
\hat P_Q, \quad \mbox{\ with\ }
\hat\gamma_Z = - \mu \frac{d}{d \mu} \hat Z \, \hat Z^{-1}
= \g \frac{\partial}{\partial \g} \hat Z^{[1]}.
\end{equation}

\subsection{Special conformal Ward identity.}

The renormalization of the  special conformal Ward identity can now be
performed along the same line as in the previous subsection for
the dilatation. For the renormalization of the product
$[{\cal O}_{jl}] \delta_-^C S$ we obtain from Eqs.\ (\ref{VarAction}),
(\ref{trace-anomaly}) together with (\ref{ren-A}), (\ref{ren-B}),
(\ref{ren-ext}) and (\ref{ren-EOM}):
\begin{eqnarray}
i [{\cal O}_{jl}] \delta_-^C S
&=& - i \sum_{k = 0}^{j}
\left\{
\frac{\beta_\epsilon}{\g} \hat Z_A^-
+ \left( \frac{\beta_\epsilon}{\g} - \gamma_G \right) \hat Z_B^-
+ 2 ( \gamma_\psi - \epsilon )\hat Z \hat b \hat P_Q \hat Z^{-1}
\right\}_{jk} [ {\cal O}_{kl-1} ]
\nonumber\\
&& - \frac{\beta_\epsilon}{\g} i [ {\cal O}_{jl}{\cal O}_A^-]
- \left( \frac{\beta_\epsilon}{\g} - \gamma_G \right)
i [ {\cal O}_{jl} {\cal O}_{B^\ast}^-]
+ \left( \gamma_G - \frac{\beta}{\g} \right )
i [{\cal O}_{jl}{\mit \Omega}^-_{G}]
\\
&&
+ ( \gamma_\psi - \epsilon )
i [{\cal O}_{jl} {\mit \Omega}^-_{\bar\psi\psi}]
+ 2 \gamma_\omega
i [{\cal O}_{jl}{\mit \Omega}^-_{\bar{\omega}}]
+ ( d - 2 ) i [ {\cal O}_{jl} \Delta^-_{\rm ext}] ,
\nonumber
\end{eqnarray}
where $[\Delta^-_{\rm ext}] = \int d^dx\, 2 x_- \partial_\mu
[ O_{B\mu} (x) ]$. Inserting this equation together with the
special conformal variation of the composite operator, given in
Eq.\ (\ref{VarOper}), into the Ward identity (\ref{CWI}) (with
$\delta\equiv\delta^C_-$) we obtain the desired CWI
\begin{eqnarray}
\label{sCWI}
\langle [ {\cal O}_{jl} ] \delta^C_- \X \rangle
&=& -i\sum_{k = 0}^{j}
\left\{ \hat a(l) + \hat\gamma^c (l) \right\}_{jk}
\langle  [ {\cal O}_{kl-1} ] \X \rangle
+ \frac{\beta}{\g}
\langle i [ {\cal O}_{jl} {\cal O}_A^- ] \X \rangle
+ \left( \frac{\beta}{\g} - \gamma_G \right)
\langle i [ {\cal O}_{jl} {\cal O}_{B^\ast}^-] \X \rangle
\nonumber\\
&&\hspace{-1.5cm}
- 2\langle i [ {\cal O}_{jl} \Delta^-_{\rm ext} ] \X \rangle
-\gamma_\psi
\langle i[ {\cal O}_{jl} {\mit \Omega}_{\bar\psi\psi}^- ] \X \rangle
-\left( \gamma_G - \frac{\beta}{\g} \right )
\langle i[ {\cal O}_{jl} {\mit \Omega}_{G}^- ] \X \rangle
- 2\gamma_\omega
\langle i[ {\cal O}_{jl} {\mit \Omega}_{\bar\omega}^- ] \X \rangle .
\end{eqnarray}
The anomaly appearing here is called special conformal anomaly
matrix and is defined by
\begin{eqnarray}
\label{def-scAno}
\hat\gamma^c(l)
&=& \lim_{\epsilon \to 0}
\left\{
\hat Z \left[
\hat a(l) - 2 ( \gamma_\psi - \epsilon ) \hat P_Q \hat b
\right]
\hat Z^{-1}
- \frac{\beta_\epsilon}{\g} \hat Z^-_A
- \left( \frac{\beta_\epsilon}{\g} - \gamma_G \right) \hat Z^-_B
- \hat a(l) \right\}
\nonumber\\
&=&
- 2 \gamma_\psi P_Q \hat b
+ 2 \left[ \hat Z^{[1]}, \hat P_Q \hat b \right]
+ \hat Z^{[1]-}_A + \hat Z^{[1]-}_B ,
\end{eqnarray}
where the second equality holds due to the finiteness of
$\hat\gamma^c$. Note that this definition looks similar to the
anomalous dimension matrix defined in Eq.\ (\ref{def-delaAno}).

\section{Commutator constraints.}

A priori there is no way to relate the new renormalization matrices
$\hat Z_{A}^-$ and $\hat Z_{B}^-$ to the renormalization group
constants. Obviously, the special conformal anomaly matrix
$\hat\gamma^c(l)$ contains therefore new information. However, from
the conformal algebra it is possible to derive the following constraints
for the conformal anomaly matrices:
\begin{eqnarray}
\label{const-1}
[ \K_-, {\cal P}_+]_- = - 2 i ( \D + {\cal M}_{-+} ) \quad
&\Rightarrow & \quad
\hat\gamma^c (l + 1) - \hat\gamma^c ( l )
= - 2 \left( \hat\gamma - 2 \frac{\beta}{\g} \hat P_G \right) ,
\\
\label{const-2}
[ \D, \K_-]_- = i \K_- \quad
&\Rightarrow & \quad
\left[
\hat\gamma ,
\hat a (l) + \hat\gamma^c (l) + 2 \frac{\beta}{\g} \hat b (l) \hat P_Q
\right]_- = 0 .
\end{eqnarray}
The constraint (\ref{const-1}) tells us that the breaking of translation
invariance by the infinitesimal special conformal transformation is
controlled by the commutator relation between ${\cal K}_-$ and
${\cal P}_+$ providing the spin dependence of the special conformal
anomaly matrix. The constraint (\ref{const-2}) relates the off-diagonal
matrix elements of $\hat\gamma$ to the matrix $\hat\gamma^c (l) + 2
\frac{\beta}{\g} \hat b (l) \hat P_Q$. Since $\hat a (l)$ is a diagonal
matrix, which is independent of the coupling $\alpha_s$, the $n$-loop
approximation of $\hat\gamma^{\rm ND}$ is determined by the
$(n-1)$-loop approximation of $\gamma^c (l)$ and $\beta$-function.

\subsection{Constraint for the special conformal anomaly matrix.}

From the definitions (\ref{confVaria},\ref{confVaria-1}) it follows
that the commutator $[\K_-, {\cal P}_+]_- = - 2 i ( \D + {\cal M}_{-+} )$
corresponds to the following relation for the variations $[\delta_+^P,
\delta_-^C ]_- = 2 ( \delta^S + \delta_{-+}^M )$. We will employ this
identity in order to prove the constraint (\ref{const-1}) which comes
from the conformal algebra and the Ward identities (\ref{WI-Poincare})
and (\ref{DWI}):
\begin{eqnarray}
\label{const-1-der0}
&&\langle
[ {\cal O}_{jl} ]
\left[\delta^P_+,\delta^C_-\right]_- \X
\rangle
= \langle [ {\cal O}_{jl} ]
\ 2 \left( \delta^S + \delta^M_{-+} \right) \X \rangle
\nonumber\\
&&\qquad =
\sum_{k = 0}^{j}
\left\{ 4( l + 2 ) \hat 1 + 2 \hat\gamma \right\}_{jk}
\langle  [ {\cal O}_{kl} ] \X \rangle
+ 2 \frac{\beta}{\g}
\langle i [ {\cal O}_{jl} {\cal O}_A] \X \rangle
+ 2 \left( \frac{\beta}{\g} - \gamma_G \right)
\langle i [ {\cal O}_{jl} {\cal O}_{B^\ast}] \X \rangle
\nonumber\\
&&\qquad - 2 \gamma_\psi
\langle i[ {\cal O}_{jl} {\mit \Omega}_{\bar\psi\psi} ] \X \rangle
-2 \left( \gamma_G - \frac{\beta}{\g} \right )
\langle i[ {\cal O}_{jl} {\mit \Omega}_{G} ] \X \rangle
- 2 \gamma_\omega
\langle i[ {\cal O}_{jl} {\mit \Omega}_{\bar\omega} ] \X \rangle ,
\end{eqnarray}
With the help of the Ward identities (\ref{WI-Poincare}) and
(\ref{sCWI}) we find immediately for the first term in the commutator
on the LHS of Eq.\ (\ref{const-1-der0}):
\begin{eqnarray}
\label{const-1-der1}
&&\langle [ {\cal O}_{jl} ] \delta^P_+ \delta^C_- \X \rangle
= - \sum_{k = 0}^{j}
\left\{ \hat a (l+1) + \hat\gamma^c (l+1) \right\}_{jk}
\langle  [ {\cal O}_{kl} ] \X \rangle
+  \frac{\beta}{\g}
\langle  [ {\cal O}_{jl+1} {\cal O}_A^- ] \X \rangle
\nonumber\\
&&\qquad + \left( \frac{\beta}{\g} - \gamma_G \right)
\langle  [ {\cal O}_{jl+1} {\cal O}_{B^\ast}^-] \X \rangle
- 2 \langle  [ {\cal O}_{jl} \Delta^-_{\rm ext} ] \X \rangle
- \gamma_\psi
\langle [ {\cal O}_{jl+1} {\mit \Omega}_{\bar\psi\psi}^- ] \X \rangle
\nonumber\\
&&\qquad - \left( \gamma_G - \frac{\beta}{\g} \right )
\langle [ {\cal O}_{jl+1} {\mit \Omega}_{G}^- ] \X \rangle
- 2 \gamma_\omega
\langle [ {\cal O}_{jl+1} {\mit \Omega}_{\bar\omega}^- ] \X \rangle.
\end{eqnarray}
While for the second term we get
\begin{eqnarray}
\label{const-1-der2}
&&\langle [ {\cal O}_{jl} ] \delta^C_- \delta^P_+ \X \rangle
= -\sum_{k = 0}^{j}
\left\{\hat a(l)+\hat\gamma^c(l) \right\}_{jk}
\langle  [ {\cal O}_{kl+1} ] \X \rangle
+  \frac{\beta}{\g}
\langle i [ {\cal O}_{jl} {\cal O}_A^- ] \delta^P_+ \X \rangle
\nonumber\\
&&\qquad + \left(\frac{\beta}{\g}-\gamma_G\right)
\langle i [ {\cal O}_{jl} {\cal O}_{B^\ast}^-] \delta^P_+ \X \rangle
- 2 \langle i [ {\cal O}_{jl} \Delta^-_{\rm ext} ] \delta^P_+ \X \rangle
-\gamma_\psi
\langle
i[ {\cal O}_{jl} {\mit \Omega}_{\bar\psi\psi}^- ]
\delta^P_+ \X \rangle
\nonumber\\
&&\qquad - \left( \gamma_G - \frac{\beta}{\g} \right )
\langle
i[ {\cal O}_{jl} {\mit \Omega}_{G}^- ]
\delta^P_+\X \rangle
- 2 \gamma_\omega
\langle
i[ {\cal O}_{jl} {\mit \Omega}_{\bar\omega}^- ]
\delta^P_+ \X \rangle.
\end{eqnarray}
To proceed further we have to get rid of the variation sign on the field
monomial, i.e.\ $\delta^P_+ \X$.

Using the definition of the subtractive renormalization
(\ref{ren-A}), (\ref{ren-B}) for the product of two composite operators
we have
\begin{eqnarray}
\label{const-1-der3}
\langle i[{\cal O}_{jl}{\cal O}^-_i ] \delta^P_+ \X \rangle
&=&
-\langle i \delta^P_+ [{\cal O}_{jl} {\cal O}^-_i] \X \rangle
\quad \mbox{for}\quad
i=\{ A, B^\ast\}
\nonumber\\
&=&
\langle [{\cal O}_{jl+1} {\cal O}^-_i] \X \rangle
- 2 \langle i [{\cal O}_{jl} {\cal O}_i] \X \rangle
- 4 \delta_{i A} \hat P_G \langle [ {\cal O}_{jl} ] \X \rangle .
\end{eqnarray}
Note that the last term arises due to the non-minimal nature of the
subtraction: the renormalization matrix $\hat Z_A$ consists of a finite
term apart from divergent contributions $\propto\frac{1}{\epsilon}$.

Since the operator $[\Delta^-_{\rm ext}] = - 2 n^\ast_\mu \int d^d x \
[ {\cal O}_{B\mu} ]$ is translation invariant we can write
\begin{eqnarray}
\label{const-1-der4}
\langle i [{\cal O}_{jl} \Delta^-_{\rm ext}] \delta^P_+ \X \rangle
= \langle  [ {\cal O}_{jl+1} \Delta^-_{\rm ext} ] \X \rangle.
\end{eqnarray}
For the terms containing the EOM we find immediately
\begin{eqnarray}
\label{const-1-der5}
\langle
i[ {\cal O}_{jl} {\mit \Omega}_{\phi}^- ]
\delta^P_+\X \rangle
\!\!\!&=&\!\!\!
2 \langle [ {\cal O}_{jl}]
\int dx\, \phi(x) \frac{\delta}{\delta\phi(x)} \X \rangle
-
\langle i[ {\cal O}_{jl}] \delta^P_+
\int dx\, 2x_- \phi(x) \frac{\delta}{\delta\phi(x)} \X \rangle
\nonumber\\
&=&\!\!\!
\langle [ {\cal O}_{jl+1} {\mit \Omega}^-_{\phi}] \X \rangle
- 2 \langle i [ {\cal O}_{jl} {\mit \Omega}_{\phi}] \X \rangle .
\end{eqnarray}
Inserting our findings (\ref{const-1-der1}-\ref{const-1-der4})
in equation (\ref{const-1-der0}) and using the definition of the
elements of the $\hat a$-matrix yields
\begin{eqnarray}
\label{const-1-der6}
\sum_{k = 0}^{j}
\left\{
\hat\gamma^c ( l + 1 )
- \hat\gamma^c ( l )
+ 2 \left( \hat\gamma - 2 \frac{\beta}{\g} \hat P_G \right)
\right\}_{jk}
\langle  [ {\cal O}_{kl+1} ] \X \rangle = 0.
\end{eqnarray}
Owing to linear independence of the Gegenbauer polynomials we can
safely omit the Green function and get finally the matrix equation
(\ref{const-1}).

\subsection{Constraint for the off-diagonal anomalous dimension matrix.}

Now we prove the commutator constraint (\ref{const-2}). Since
$\delta^S = - \mu \frac{\partial}{\partial\mu}$, instead of dealing
with the corresponding commutator we apply the renormalization group
operator
\begin{eqnarray*}
D = \mu \frac{d}{d\mu} \hat 1 + \hat\gamma
\end{eqnarray*}
on both sides of the special CWI (\ref{sCWI}) without taking into
account the differentiation of the field monomial $\X$ with respect
to $\mu$ since these terms vanish by means of the special conformal
Ward identity. The LHS vanishes identically since $D [{\cal O}_{jk}]
= 0$. To calculate the RHS we employ formally the renormalization
group equations of the appearing operator insertions. For this purpose
we need the following commutators
\begin{equation}
\left[
\mu \frac{d}{d \mu}, \g \frac{\partial}{\partial\g}
\right]_-
= - \left(
\g \frac{\partial}{\partial\g} \frac{\beta_\epsilon}{\g}
\right)
\g \frac{\partial}{\partial\g}
+ 2 \left(
\g \frac{\partial}{\partial\g} \gamma_G
\right)
\xi \frac{\partial}{\partial\xi} , \ \
\left[
\mu \frac{d}{d \mu}, \xi \frac{\partial}{\partial\xi}
\right]_-
= 2 \left(
\xi \frac{\partial}{\partial\xi} \gamma_G
\right)
\xi \frac{\partial}{\partial\xi} .
\end{equation}
Using them properly when one acts with the differential operator $\mu
\frac{d}{d\mu}$ on the operator insertions  $[{\cal O}_i^-]$,
we obtain the desired renormalization group equations:
\begin{eqnarray}
\mu \frac{d}{d\mu} [{\cal O}_A^-]
&=&
- \left(
\g \frac{\partial}{\partial\g} \frac{\beta}{\g}
\right)
[{\cal O}_A^-]
-
\left(
\g \frac{\partial}{\partial\g}
-
2 \xi \frac{\partial}{\partial\xi}
\right)
\left(
\frac{\beta}{\g} - \gamma_G
\right)
[{\cal O}_{B^\ast}^-] \\
&&\hspace{5cm}+
\sum_{\phi}
\left(
\g \frac{\partial}{\partial\g}
-
2 \xi \frac{\partial}{\partial\xi}
\right) {\cal F}_\phi(\g)
{\mit \Omega}_\phi^- ,
\nonumber\\
\mu \frac{d}{d\mu} [{\cal O}_{B^\ast}^-]
&=&
\left(
2 \xi \frac{\partial}{\partial\xi} \gamma_G
\right)
[{\cal O}_{B^\ast}^-]
+
\sum_{\phi}
\left(
2 \xi \frac{\partial}{\partial\xi} {\cal F}_\phi(\g)
\right)
{\mit \Omega}_\phi^- ,
\end{eqnarray}
where
\begin{eqnarray*}
{\cal F}_G (\g) = \gamma_G - \frac{\beta}{\g}, \qquad
{\cal F}_{\bar\psi\psi} (\g) = \gamma_\psi, \qquad
{\cal F}_{\bar\omega} (\g) = 2 \gamma_\omega .
\end{eqnarray*}

Now we list the action of the operator $D$ on every term appearing
on the RHS of the special conformal Ward identity. The first equation
is obvious:
\begin{eqnarray}
D \sum_{k = 0}^{j}
\left\{ \hat a (l) + \hat \gamma^c (l) \right\}_{jk}
\langle [ {\cal O}_{kl-1} ] \X \rangle
\!\!\!\!&=&\!\!\!\! \sum_{k = 0}^{j}
\left\{
\left[
\hat\gamma , \hat a + \hat\gamma^c + 2\frac{\beta}{\g} \hat P_Q \hat b
\right]_- \right. \\
\!\!\!\!&+&\!\!\!\!
\left.
\beta \frac{\partial}{\partial\g}
\left[
\hat Z^{[1]-}_A
+
\hat Z^{[1]-}_B
-
2 \gamma_\psi \hat P_Q \hat b
\right]
\right\}_{jk}
\langle [ {\cal O}_{kl-1} ] \X \rangle . \nonumber
\end{eqnarray}
Here we have taken into account that the special conformal anomaly
matrix is a gauge independent quantity. Next we calculate the action
of $D$ on the renormalized operator products by using the subtractive
renormalization prescriptions (\ref{ren-A}) and (\ref{ren-B}) as well
as the renormalization group equations derived above:
\begin{eqnarray}
&&\hspace{-0.5cm}D \frac{\beta}{\g}
\langle [ {\cal O}_{jl} {\cal O}_A^- ] \X \rangle \\
&&\quad =
\frac{\beta}{\g}
\left(
\g \frac{\partial}{\partial\g}
-
2 \xi \frac{\partial}{\partial\xi}
\right)
\left(
\gamma_G - \frac{\beta}{\g}
\right)
\langle [ {\cal O}_{jl} {\cal O}_{B^\ast}^- ] \X \rangle
+
\frac{\beta}{\g}
\sum_{\phi}
\left(
\g \frac{\partial}{\partial\g}
-
2 \xi \frac{\partial}{\partial\xi}
\right)
{\cal F}_\phi(\g)
\langle [ {\cal O}_{jl} {\mit \Omega}_\phi^- ] \X \rangle \nonumber\\
&&\quad +
\frac{\beta}{\g}
\sum_{k = 0}^{j}
\left\{
\left( \g \frac{\partial}{\partial\g} \hat Z^{[1]-}_A \right)
- 2 \left(
\g \frac{\partial}{\partial\g}
-
2 \xi \frac{\partial}{\partial\xi}
\right)
\gamma_\psi \hat P_Q \hat b
\right\}_{jk}
\langle [ {\cal O}_{kl - 1} ] \X \rangle
, \nonumber\\
&&\hspace{-0.5cm}D \left( \frac{\beta}{\g} - \gamma_G \right)
\langle [ {\cal O}_{jl} {\cal O}_{B^\ast}^- ] \X \rangle \\
&&\quad =
- \frac{\beta}{\g}
\left(
\g \frac{\partial}{\partial\g}
-
2 \xi \frac{\partial}{\partial\xi}
\right)
\left(
\gamma_G - \frac{\beta}{\g}
\right)
\langle [ {\cal O}_{jl} {\cal O}_{B^\ast}^- ] \X \rangle
+
\left( \frac{\beta}{\g} - \gamma_G \right)
\sum_{\phi}
\left(
2 \xi \frac{\partial}{\partial\xi}
{\cal F}_\phi(\g)
\right)
\langle [ {\cal O}_{jl} {\mit \Omega}_\phi^- ] \X \rangle \nonumber\\
&&\quad +
\left( \frac{\beta}{\g} - \gamma_G \right)
\sum_{k = 0}^{j}
\left\{
\left( \g \frac{\partial}{\partial\g} \hat Z^{[1]-}_B \right)
- 4 \left(
\xi \frac{\partial}{\partial\xi} \gamma_\psi
\right)
\hat P_Q \hat b
\right\}_{jk}
\langle [ {\cal O}_{kl - 1} ] \X \rangle .
\nonumber
\end{eqnarray}
Due to Eq.\ (\ref{ren-EOM-1}) the following relation holds
\begin{equation}
D\ {\cal F}_\phi(\g)
\langle [ {\cal O}_{jl} {\mit \Omega}_\phi^- ] \X \rangle
=
\left(
\mu \frac{d}{d \mu} {\cal F}_\phi(\g)
\right)
\langle [ {\cal O}_{jl} {\mit \Omega}_\phi^- ] \X \rangle .
\end{equation}
In the above equations we have omitted the divergent contributions
on the RHS (which cancel each other) since the LHS is finite. All
other terms in the special conformal Ward identity vanish when the
$D$-operator acts on them.

Collecting all terms listed above, we find the desired constraint
\begin{eqnarray}
\label{constraint}
\sum_{k=0}^{j}\left\{\left[
\hat\gamma ,
\hat a (l) + \hat\gamma^c (l) + 2 \frac{\beta}{\g} \hat b (l) \hat P_Q
\right]_-\right\}_{jk} \langle [ {\cal O}_{kl-1}] \X \rangle  = 0,
\end{eqnarray}
which implies finally  the matrix equation (\ref{const-2}).

We have also found the following expression for the gauge-dependent
renormalization constant $\hat Z_B^-$:
\begin{equation}
\label{Z-b-constant}
\g \frac{\partial}{\partial\g} \hat Z_B^{[1]-}
=
4 \left( \xi \frac{\partial}{\partial\xi} \gamma_\psi \right)
\hat P_Q \hat b .
\end{equation}

\section{Leading order conformal anomalies.}

In this section we compute the conformal anomalies to LO in a general
covariant gauge. Since we are interested in the present paper in the NLO
approximation for the anomalous dimension matrix of conformal
operators, we restrict ourselves to the LO approximation of the special
conformal anomaly which reads
\begin{equation}
\hat\gamma^c
=
\left(
\begin{array}{ll}
2 \left[ {^{QQ}\!\hat Z^{[1]}} , \hat b \right]_-
+ {^{QQ}\!\hat Z_A^{[1]-}}
&\ - 2\, \hat b\, {^{QG}\!\hat Z^{[1]}}
\\
2\, {^{GQ}\!\hat Z^{[1]}}\, \hat b
+ {^{GQ}\!\hat Z_A^{[1]-}}
&\ {^{GG}\!\hat Z_A^{[1]-}}
\end{array}
\right) .
\end{equation}
Here we have taken into account that in one-loop approximation the
following equalities hold
\begin{equation}
{^{QQ}\!\hat Z_B^{[1]-}} = 2 \gamma_\psi \hat b
\quad \mbox{and} \quad
{^{QG}\!\hat Z_A^{[1]-}} = 0.
\end{equation}

For technical reason we perform the calculation in the light-cone position
formalism. Then it is quite straightforward to transform these results
to any other representation. We define the relevant light-ray operators as
follows:
\begin{eqnarray}
{\cal O}(\kappa_1,\kappa_2)=
\left(
{ {^Q\!{\cal O}(\kappa_1,\kappa_2)}
\atop
{^G\!{\cal O}(\kappa_1,\kappa_2)}}
\right),
\end{eqnarray}
where
\begin{eqnarray}
\label{treeQLRO}
\left\{\!\!\!
\begin{array}{c}
{^Q\!{\cal O}^V} \\
{^Q\!{\cal O}^A}
\end{array}
\!\!\!\right\}
(\kappa_1,\kappa_2)\!
\!\!&=&\!\!
\bar{\psi}(\kappa_2 n)
\!\left\{\!\!\!
\begin{array}{c}
\gamma_+ \\
\gamma_+ \gamma_5
\end{array}\!\!\!\right\}
\Phi [\kappa_2 n, \kappa_1 n]
\psi (\kappa_1 n)
\mp
\bar{\psi}(\kappa_1 n)
\!\left\{\!\!\!
\begin{array}{c}
\gamma_+ \\
\gamma_+ \gamma_5
\end{array}\!\!\!\right\}
\Phi [\kappa_1 n, \kappa_2 n]
\psi(\kappa_2 n)
, \\
\label{treeGLRO}
\left\{\!\!\!
\begin{array}{c}
{^G\!{\cal O}^V} \\
{^G\!{\cal O}^A}
\end{array}
\!\!\!\right\}
(\kappa_1,\kappa_2)\!
\!\!&=&\!\!
G_{+ \mu} (\kappa_2 n)
\!\left\{\!\!\!
\begin{array}{c}
g_{\mu\nu} \\
i\epsilon_{\mu\nu-+}
\end{array}\!\!\!\right\}
\Phi [\kappa_2 n, \kappa_1 n]
G_{\nu +} (\kappa_1 n).
\end{eqnarray}
The path-ordered link factor $\Phi [x_2, x_1]$ ensures gauge
invariance and reads
\begin{equation}
\Phi [x_2 , x_1]
=P \exp
\left(
i \mu^\epsilon \g X ( x_2 - x_1 )_\mu \int_{0}^{1}\! d \tau
T^a B^a_\mu \bigl( x (\tau) \bigr)
\right) ,
\end{equation}
where the path is parametrized as $ x (\tau) = \tau x_2 + \bar\tau x_1$.
The local conformal composite operators (\ref{treeCO}) are obtained by
differentiation with respect to $\kappa_1$ and $\kappa_2$, namely:
\begin{equation}
{\cal O}_{jl}=
\left.\left(
{\hspace{5mm}
\frac{1}{2}( i \partial_+ )^l C^{\frac{3}{2}}_j \!
\left( \frac{\stackrel{\leftrightarrow}{\partial_+}}{\partial_+} \right)
{ {^Q\!{\cal O}(\kappa_1,\kappa_2)}}
\atop
{( i \partial_+ )^{l-1} C^{\frac{5}{2}}_{j-1}\!
\left( \frac{\stackrel{\leftrightarrow}{\partial_+}}{\partial_+} \right)
{^G\!{\cal O}(\kappa_1,\kappa_2)}}}
\right) \right|_{\kappa_1 = \kappa_2 = 0},
\end{equation}
where $\partial_+ = \partial_{\kappa_1} + \partial_{\kappa_2}$ and
$\stackrel{\leftrightarrow}{\partial_+} = \partial_{\kappa_1} -
\partial_{\kappa_2}$.

In the next subsection we derive the special conformal anomaly matrix
for the quark sector including chiral-odd (transversity) twist-2 operators.
This matrix turns out to be universal in a sense that it is
the same for all Dirac structures. Then we comment shortly our results
for the mixed $QG$ and $GQ$ channels, which were derived previously in
Abelian theory. In the last subsection we evaluate the special conformal
anomaly matrix for the $GG$ channel for parity even and odd operators.
Again in all channels we observe the universality of these kernels, which
implies that the additional special conformal symmetry breaking is the same
in vector and axial-vector channels: the difference arises from the
different scale anomalies, i.e.\ the one-loop anomalous dimensions of
composite operators.

\subsection{Once more about quarks: chiral-even and -odd sectors.}


\begin{figure}[t]
\begin{center}
\vspace{5.0cm}
\hspace{-4cm}
\mbox{
\begin{picture}(0,220)(270,0)
\put(0,-30)                    {
\epsffile{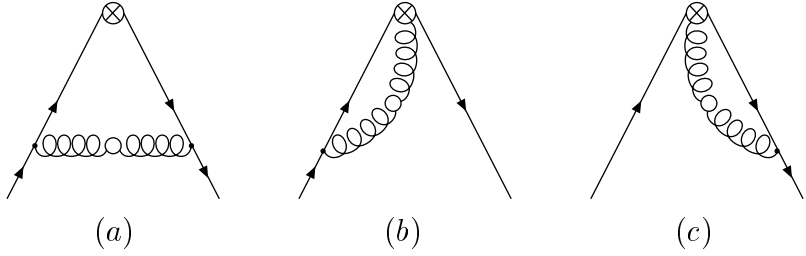}
                               }
\end{picture}
}
\end{center}
\vspace{-9.4cm}
\caption{\label{quarks} One-loop graphs contributing to the
renormalization constant $\hat Z_A^-$.}
\end{figure}

Although the quark sector was considered by us in detail in
previous studies, nevertheless, it is worthwhile to address this issue
from our present point of view. Moreover, we complete the list of
two-loop anomalous dimensions by adding those of the chiral-odd
(transversity) sector. Thus we should evaluate the renormalization constant
${^{QQ}\!\hat Z_A^{[1]-}}$ from diagrams in Fig.\ \ref{quarks}.
This calculation is quite trivial. Using the Feynman rules given in
appendix \ref{FR} we get easily\footnote{The last argument of the
kernel stands for the value of the gauge fixing parameter.}:
\begin{eqnarray}
&&\left.
i[{\cal O}^-_A]
[{^Q\!{\cal O}^{\mit\Gamma}} (\kappa_1 , \kappa_2)]
\right|_{(a)}
=
\frac{\alpha_s}{2\pi}
\frac{i}{\epsilon}
\int_{0}^{1} dz \int_{0}^{\bar z} dy \\
&&\qquad\qquad\qquad\times
2\, {^{QQ}\!\K^{\mit\Gamma}_{(a)}} (y, z | \xi = 0 )
\left( \partial^{k_1}_- + \partial^{k_2}_- \right)
[{^Q\!{\cal O}^{\mit\Gamma}}
( \bar y \kappa_1 + y \kappa_2 , z \kappa_1 + \bar z \kappa_2 )],
\nonumber\\
&&\left.
i[{\cal O}^-_A]
[{^Q\!{\cal O}^{\mit\Gamma}} (\kappa_1 , \kappa_2)]
\right|_{(b + c)}
=
\frac{\alpha_s}{2\pi}
\frac{i}{\epsilon}
\int_{0}^{1} dz \int_{0}^{\bar z} dy \\
&&\qquad\qquad\qquad\times
\biggl\{
\! 2\, {^{QQ}\!\K^{\mit\Gamma}_{(b + c)}} (y, z | \xi = 0 )
\left( \partial^{k_1}_- + \partial^{k_2}_- \right)
+
{^{QQ}\!\K^w} (y, z)
\biggr\}
[{^Q\!{\cal O}^{\mit\Gamma}}
( \bar y \kappa_1 + y \kappa_2 , z \kappa_1 + \bar z \kappa_2 )] .
\nonumber
\end{eqnarray}
The separate contributions ${^{QQ}\!\K^{\mit\Gamma}_i}$ to the
evolution kernel correspond to the Landau gauge, so that the sum
${^{QQ}\!\K^{\mit\Gamma}} = {^{QQ}\!\K^{\mit\Gamma}_{(a)}} +
{^{QQ}\!\K^{\mit\Gamma}_{(b+c)}}$ is gauge invariant as it should
be, since for $\xi = 0$ the one-loop quark anomalous dimension
vanishes, $\gamma_\psi^{(0)} (\xi = 0) = 0$. The $\K^w$-kernel reads
\begin{equation}
\label{W-kernel}
{^{QQ}\!\K^w} (y, z) = -2 C_F
\left\{
\frac{1}{k_{1+}} \left[ \frac{\bar y}{y^2} \right]_+ \delta (z)
+
\frac{1}{k_{2+}} \left[ \frac{\bar z}{z^2} \right]_+ \delta (y)
\right\},
\end{equation}
where we have introduced the plus-prescription for a singularities $1/y^k$
(with $k = 1,2,\dots$)
\begin{equation}
\left[ \frac{1}{y^k} \right]_+
= \frac{1}{y^k}
- \sum_{j = 0}^{k - 1} \frac{(-1)^j}{j!}
\delta^{(j)} (y) \int_0^1 \frac{dz}{z^{k - j}}.
\end{equation}
When Eq.\ (\ref{W-kernel}) is transformed into the light-cone fraction
representation (see Ref.\ \cite{BelMue97b}) we reproduce the known
expression for the $w (x,y)$-function \cite{Mue94}:
\begin{equation}
{^{QQ}\!w} (x, y) = -2 C_F
\left[
\frac{x}{y} \frac{\theta (y - x)}{(x - y)^2}
+
\left(
{ x \to \bar x \atop y \to \bar y }
\right)
\right]_+ .
\end{equation}
This kernel is the same for all quark operators irrespective of
their Dirac structure: ${\mit\Gamma}= \gamma_+, \ \gamma_+\gamma_5$ or
$\sigma_{+\perp}\gamma_5$. Therefore, the special conformal
anomaly matrix reads ${^{QQ}\!\hat\gamma^{c(0)}} = - \hat b \,
{^{QQ}\!\hat\gamma^{(0)}} + {^{QQ}\!\hat w}$.

\subsection{Mixed channels.}

Note, that in the present approach there is no need to perform any
calculation for the mixing $QG$ channel. The $w$-kernel for the
$GQ$-channel was calculated by us in Ref.\ \cite{BelMue98} with the same
result for the parity even and odd channels
\begin{equation}
{^{GQ}\cal K}^w (y, z) = C_F
[ \delta (z) - \delta (y) ] .
\end{equation}
The momentum space kernel reads
\begin{equation}
{^{GQ}\!w} (x, y) = 2 C_F
\left\{
\frac{1}{y} \theta (y - x)
-
\left( {x \to \bar x \atop y \to \bar y } \right)
\right\} .
\end{equation}
Note the extra factor of 2 in front of the light-cone fraction kernel
which comes as a result of the transformation from the coordinate to the
momentum space according to Eqs.\ (\ref{NFPD-Q},\ref{NFPD-G}).

\subsection{Gluonic sector.}

Using the representation of the conformal operators in terms of the
derivatives of non-local string operators we can write the
divergent part of the operator product $i [ {\cal O}_A^- ]
[ {^G\!{\cal O}( \kappa_1, \kappa_2 )} ]$ via the following relation
in leading order of the coupling constant:
\begin{eqnarray}
i [ {\cal O}_A^- ] [ {^G\!{\cal O}( \kappa_1, \kappa_2 )} ]
&=& \frac{\alpha_s}{2 \pi} \frac{i}{\epsilon}
\int_{0}^{1} dz \int_{0}^{\bar z} dy
\biggl\{
{^{GG}\!\K_A^-} ( y, z )
[ {^G\!{\cal O}( \bar y \kappa_1 + y \kappa_2 ,
z \kappa_1 + \bar z \kappa_2 )} ] \nonumber\\
&+&
{^{GG}\!\widetilde\K_A^-} ( y, z )
\int d^d x \, 2 x_- B_\mu^b (x)
\frac{\delta}{\delta B_\mu^b (x) }
[ {^G\!{\cal O}( \bar y \kappa_1 + y \kappa_2 ,
z \kappa_1 + \bar z \kappa_2 )} ]
\biggr\}.
\end{eqnarray}
Here the kernels $\K (y, z)$ are the subjects of the calculations.
The diagrammatic representation of the corresponding graphs is given
in Fig.\ \ref{one-loop} and the corresponding Feynman rules can be found
in Appendix \ref{FR}. The details of the calculations together
with the general approach to the evolution equations for non-local
string operators are given in Appendix \ref{renormalization}.


\begin{figure}[t]
\begin{center}
\vspace{4.8cm}
\hspace{-2.5cm}
\mbox{
\begin{picture}(0,220)(270,0)
\put(0,-30)                    {
\epsffile{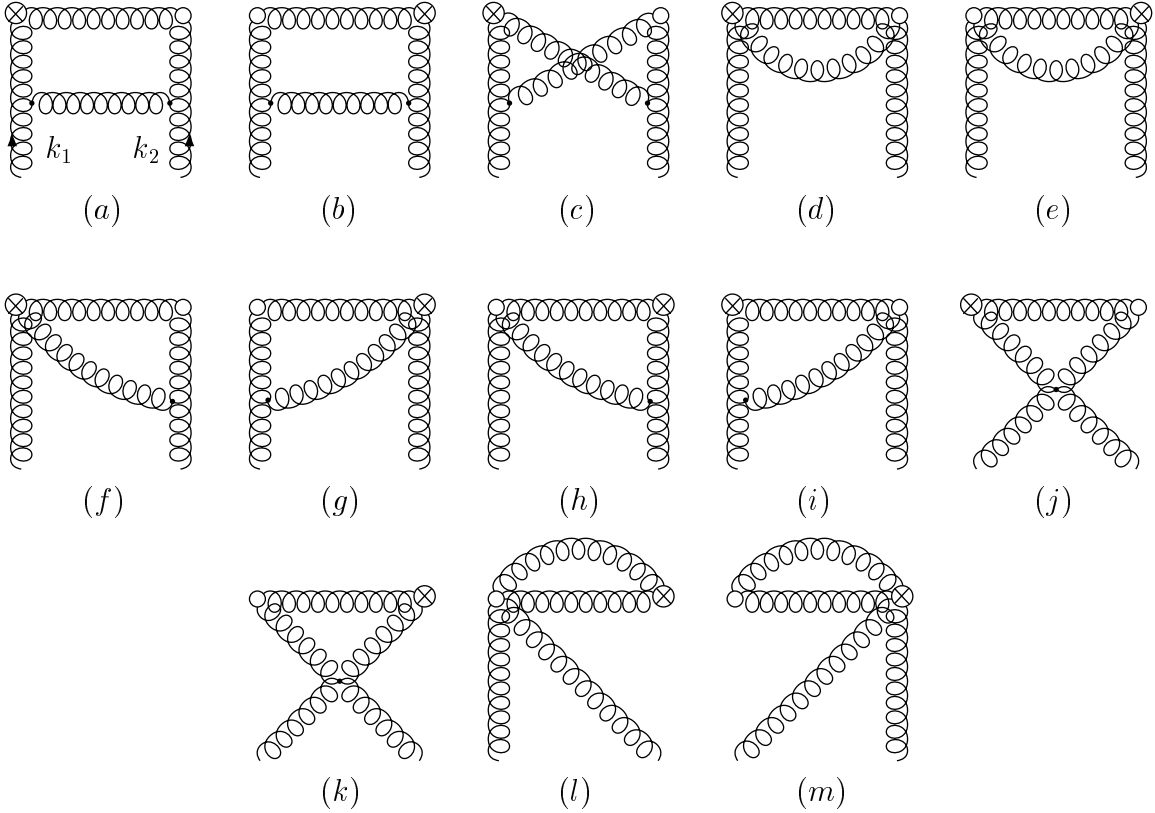}
                               }
\end{picture}
}
\end{center}
\vspace{-2cm}
\caption{ \label{one-loop} One-loop diagrams which give rise to
divergencies in the product of the renormalized operator insertions
$i [ {\cal O}_A^- ][ {^G\!{\cal O}_{jl}} ]$. The symmetry factors
$s_\sigma$ for the diagrams are the following: $s_\sigma = 1$ for
$\sigma = a,b,c,f,g,h,i$ and $s_\sigma = \frac{1}{2}$ for $\sigma
= d,e,j,k,l$. The diagram ($m$) is identically zero in dimensional
regularization.}
\end{figure}

Note that the factor in front of the operator $B\frac{\delta}{\delta B}
[{\cal O}]$ is gauge independent in leading order. It can be conveniently
represented as
\begin{equation}
\left(
\g \frac{\partial\ln X}{\partial\g}
- 2 \xi \frac{\partial\ln X}{\partial\xi}
\right)
=
\frac{1}{\epsilon} \frac{\alpha_s}{2\pi}
\left(
\frac{\beta_0}{2} - \gamma_G^{(0)} (\xi = 0)
\right).
\end{equation}

The result of the calculations given in detail in Appendix
\ref{renormalization} can be summarized as follows
\begin{eqnarray}
\label{gluon-ZA}
{^{GG}\!\K_A^-} ( y, z )
&=& - i 2 ( \kappa_1 + \kappa_2 ) {^{GG}\!\K} ( y, z )
- i ( \kappa_1 + \kappa_2 ) \beta_0 \delta (y) \delta (z)
+ {^{GG}\!\K^w} ( y, z ),
\\
{^{GG}\!\widetilde\K_A^-} ( y, z )
&=& i
\left(
\frac{\beta_0}{2} - \gamma_G^{(0)} (\xi = 0)
\right) \delta (y) \delta (z) ,
\end{eqnarray}
where the parity-even and -odd gluonic kernels read
\cite{GeyRob84,BalBra89,Rad96,Blum97}
\begin{eqnarray}
\label{GGV-kernel}
{^{GG}\!\K^V} ( y, z ) &=&
C_A
\biggl\{
4 ( 1 - y - z + 3 y z )
+ \left\{
y - 2 + \left[\frac{1}{y} \right]_+
\right\} \delta (z) \\
&&\hspace{4.2cm}+
\left\{
z - 2 + \left[\frac{1}{z} \right]_+
\right\} \delta (y)
\biggr\}
- \frac{\beta_0}{2} \delta (y) \delta (z), \nonumber\\
\label{GGA-kernel}
{^{GG}\!\K^A} ( y, z ) &=&
{^{GG}\!\K^V} ( y, z ) - 12 C_A y z ,
\end{eqnarray}
and
\begin{equation}
{^{GG}\!\K^w} ( y, z )
= \frac{C_A}{k_{2+}}
\left\{
4 \left[ \frac{1}{z} \right]_+
-
2 \left[ \frac{1}{z^2} \right]_+
\right\} \delta (y)
+ \frac{C_A}{k_{1+}}
\left\{
4 \left[ \frac{1}{y} \right]_+
-
2 \left[ \frac{1}{y^2} \right]_+
\right\} \delta (z) ,
\end{equation}
which is the same in both channels. The light-cone fraction function
reads from this
\begin{equation}
\label{GG-w}
{^{GG}\!w} (x, y)
= - 2 C_A
\left\{
\left[ \frac{x^2}{y^2} \frac{\theta ( y - x )}{( x - y )^2} \right]_+
- \frac{1}{y^2} \theta ( y - x )
+ \frac{1}{y} \delta ( y - x )
+ \left( { x \to \bar x \atop y \to \bar y } \right)
\right\}.
\end{equation}
Using the identity
\begin{equation}
\left.
(i \partial_+)^{l}
C^\nu_{j}
\Bigl( {\stackrel{\leftrightarrow}{\partial}_+} / {\partial_+} \Bigr)
\left\{
( \kappa_1 + \kappa_2 )
{\cal O} (\kappa_1, \kappa_2)
\right\}
\right|_{\kappa_i = 0}
=
i \sum_{k = 0}^{j} b_{jk} (l, \nu) {\cal O}_{j l - 1} ,
\end{equation}
with the $\hat b$-matrix being given by Eq.\ (\ref{b-matrix}), we can
represent the above result (\ref{gluon-ZA}) in the usual matrix form
\begin{equation}
\label{SCA-GGsector}
{^{GG}\!\hat\gamma^{c(0)}} = - \hat b \, {^{GG}\!\hat\gamma^{(0)}}
+ \beta_0\, \hat b + {^{GG}\!\hat w} ,
\end{equation}
where the $\hat w$-matrix possesses elements defined as Gegenbauer
moments of the kernel (\ref{GG-w}). This quantity will be evaluated in
the next section. Note also that Eq.\ (\ref{SCA-GGsector}) as well as
other conformal anomalies fulfill the constraint equality (\ref{const-1}).

\section{Gegenbauer moments of the evolution kernels.}

This section is mainly technical so that rather experienced reader
can skip this presentation. Here we intend to give the method
for evaluation of the moments of the evolution kernels we have derived
in the preceding section in the basis of Gegenbauer polynomials.
We deal here only with gluonic kernels.

Let us demonstrate the main technical steps for the integration of
the functions with $+$-prescription, e.g.\ the first term in Eq.\
(\ref{GG-w}), ${^{GG}\!w^{(1)}} (x, y) = [ x^2/y^2 \theta ( y - x )
/ ( x - y )^2 ]_+ + ( x \to \bar x, y \to  \bar y )$ . Integrating by
parts over $x$ we decrease the power of the singularity and end up
with equation (according to Eq.\ (\ref{CtoF}) we have replaced the
Gegenbauer polynomial by ${_2F_1}$)
\begin{equation}
{^{GG}\!w^{(1)}_{jk}}
=
- [ 1 + ( - 1)^{j - k} ] \frac{\Gamma (j + 4) \Gamma (k + 4) }{
\Gamma^2 (5) \Gamma (j) \Gamma (k) }
\int_{0}^{1} dy \frac{w (y | \frac{5}{2})}{ N_{k - 1} (\frac{5}{2})}
{_2 F_1}
\left( \left. { - k + 1 , k + 4
\atop 3 }
\right| y \right) \left\{ J_1 (y) + 2 J_2 (y) \right\} ,
\end{equation}
where the integrals $J_i$ read
\begin{eqnarray}
J_1 (y) &=& \int_{0}^{1} dx
\left[ \frac{x^2}{1 - x} \right]_+ \frac{d}{d (xy)}
{_2 F_1}
\left( \left. { - j + 1 , j + 4
\atop 3 }
\right| xy \right) , \nonumber\\
J_2 (y) &=& \int_{0}^{1} dx
x \left[ \frac{x}{1 - x} \right]_+
\frac{1}{xy}
\left[
{_2 F_1}
\left( \left. { - j + 1 , j + 4
\atop 3 }
\right| xy \right)
-
{_2 F_1}
\left( \left. { - j + 1 , j + 4
\atop 3 }
\right| 0 \right)
\right] .
\label{J-i}
\end{eqnarray}
Since at the end we get the result in terms of hypergeometric
functions and their derivatives with respect to the lower index,
the simplicity of reduction to elementary functions depends on the
handling of the integrals $J_i$. Therefore, at this step it is
instructive to expand the integrand of $J_i$ with respect to a
complete set of Gegenbauer polynomials. Namely,
\begin{eqnarray}
&&\!\!\!\!\!\!\frac{d}{d x}
{_2 F_1}
\left( \left. { - j + 1 , j + 4
\atop 3 }
\right| x \right)
=
\sum_{l = 1}^{\infty}
e_{jl}\,
{_2 F_1}
\left( \left. { - l + 1 , l + 4
\atop 3 }
\right| x \right)
, \\
&&\!\!\!\!\!\!\frac{1}{x}
\left[
{_2 F_1}
\left( \left. { - j + 1 , j + 4
\atop 3 }
\right| x \right)
-
{_2 F_1}
\left( \left. { - j + 1 , j + 4
\atop 3 }
\right| 0 \right)
\right]
=
\sum_{l = 1}^{\infty}
f_{jl}\,
{_2 F_1}
\left( \left. { - l + 1 , l + 4
\atop 3 }
\right| x \right)
.
\end{eqnarray}
The integrals (\ref{J-i}) can be easily evaluated with the result
\begin{equation}
J_1 = \sum_{l = 1}^{\infty} e_{jl}\,
\left. \frac{\partial}{\partial\epsilon} \right|_{\epsilon = 0}
\!\!\!{_2 F_1}
\left( \left. { - l + 1 , l + 4
\atop 3 + \epsilon }
\right| y \right) , \\
J_2 = \sum_{l = 1}^{\infty} f_{jl}\,
\left. \left\{
\frac{\partial}{\partial\epsilon}
- \frac{1}{2}
\right\} \right|_{\epsilon = 0}
\!\!\! {_2 F_1}
\left( \left. { - l + 1 , l + 4
\atop 3 + \epsilon }
\right| y \right) ,
\end{equation}
where we have used analytical regularization of the singular
distribution
\begin{equation}
\int_{0}^{1} \frac{dx}{[1 - x]_+} {\cal F} (x)
\stackrel{\rm Reg}{=}
\int_{0}^{1} \frac{dx}{(1 - x)^{1 - \epsilon}}
\left\{ {\cal F} (x) - {\cal F} (1) \right\} .
\end{equation}
Thus we get finally for ${^{GG}\!w^{(1)}_{jk}}$
\begin{equation}
{^{GG}\!w^{(1)}_{jk}}
= [ 1 + ( - 1)^{j - k} ]
\sum_{l = 1}^{\infty}
\left\{
f_{jl} h_{lk} (j) - ( 2 f_{jl} + e_{jl}) g_{lk} (j)
\right\},
\end{equation}
with
\begin{eqnarray}
&&\left\{
{
h_{lk} (j)
\atop
g_{lk} (j)
}
\right\}
=
\frac{\Gamma (j + 4) \Gamma (k + 4) }{
\Gamma^2 (5) \Gamma (j) \Gamma (k) }
\int_{0}^{1} dy \frac{w (y | \frac{5}{2})}{ N_{k - 1} (\frac{5}{2})}
{_2 F_1}
\left( \left. { - k + 1 , k + 4
\atop 3 }
\right| y \right) \nonumber\\
&&\hspace{9cm}\times
\left\{
\left. { 1 \atop \frac{\partial}{\partial\epsilon} }
\right\} \right|_{\epsilon = 0}
{_2 F_1}
\left( \left. { - l + 1 , l + 4
\atop 3 + \epsilon }
\right| y \right) .
\end{eqnarray}
Now these matrices can be easily computed. The main steps are reduced to
the use of the representation (\ref{C-diff}) and integration by parts
until all derivatives will act on the right hypergeometric function. Then
with the help of Eqs.\ (\ref{F-diff}) and (\ref{B-transformation}), we
obtain the result in terms of hypergeometric functions of argument 1.
The reduction to elementary functions is accomplished according to
the formulae (\ref{reduce}-\ref{deriv3F2}). In this way we get
\begin{eqnarray}
e_{jl} &=& - [ 1 - ( - 1)^{j - l} ]
\theta_{j - 1, l} (3 + 2 l) \frac{\Gamma (j)\Gamma (l + 4)}{
\Gamma (j + 4)\Gamma (l)} , \\
f_{jl} &=& - \frac{1}{2} [ 1 - ( - 1)^{j - l} ]
\theta_{j - 1, l} (3 + 2 l) \frac{\Gamma (j)\Gamma (l + 4)}{
\Gamma (j + 4)\Gamma (l)}
\left[
\frac{\Gamma (j + 4)\Gamma (l)}{\Gamma (j)\Gamma (l + 4)} - 1
\right] , \\
h_{lk} (j) &=& \delta_{lk}
\frac{\Gamma (j + 4)\Gamma (l)}{\Gamma (j)\Gamma (l + 4)} , \\
g_{lk} (j) &=& \theta_{l,k}
\frac{\Gamma (j + 4)\Gamma (l)}{\Gamma (j)\Gamma (l + 4)}
\left\{
( \psi (3) - \psi (k + 2) ) \delta_{lk}
+ [ 1 - ( - 1)^{l - k} ] \frac{(3 + 2k)}{(l - k)(l + k + 3)}
\right\} .
\end{eqnarray}
Combining these results together we have
\begin{eqnarray}
{^{GG}\!w^{(1)}_{jk}}
= [ 1 + ( - 1)^{j - k} ]\!\!\!\!\!\!&&\!\!\!\!\!\!\theta_{j - 2,k}
( 3 + 2k )
\biggl\{
( \psi (2) - \psi (k + 2) )
\left[
\frac{\Gamma (j + 4)\Gamma (k)}{\Gamma (j)\Gamma (k + 4)} - 1
\right] \nonumber\\
&+& \sum_{l = k + 1}^{j - 1}
[ 1 - ( - 1)^{l - k} ]
\frac{(3 + 2l)}{(l - k)(l + k + 3)}
\left[
\frac{\Gamma (j + 4)\Gamma (k)}{\Gamma (j)\Gamma (k + 4)} + 1
\right]
\biggr\}.
\end{eqnarray}
The final summation can be easily performed with help of the formulae
\begin{eqnarray}
&&\sum_{l = k + 1}^{j - 1}
[ 1 - ( - 1)^{l - k} ]
\frac{(3 + 2l)}{(l - k)(l + k + 3)}
= A_{jk} + \psi (j + 2) - \psi (k + 2) , \\
&&\sum_{l = k + 1}^{j - 1}
[ 1 - ( - 1)^{l - k} ]
\frac{(3 + 2l)}{(l - k)(l + k + 3)}
\frac{\Gamma (l)}{\Gamma (l + 4)}
= \frac{\Gamma (k)}{\Gamma (k + 4)}
\biggl\{ A_{jk} - \psi (j + 2) + \psi (k + 2) \nonumber\\
&&\hspace{10.5cm} - \frac{(j - k)(j + k + 3)}{(j + 1)(j + 2)}
\biggr\} .
\end{eqnarray}

Adding the contributions from the last two terms in Eq.\ (\ref{GG-w})
\begin{eqnarray}
{^{GG}\!w^{(2)}_{jk}}
\!\!\!&=&\!\!\! [ 1 + ( - 1)^{j - k} ]\theta_{j - 2,k}
( 3 + 2k )
\frac{1}{2} \frac{\Gamma (k)}{\Gamma (k + 4)}
(j - k)(j + k + 3) \left[ (j - k)(j + k + 3) + 2 \right], \\
{^{GG}\!w^{(3)}_{jk}}
\!\!\!&=&\!\!\! - [ 1 + ( - 1)^{j - k} ]\theta_{j - 2,k}
( 3 + 2k )
\frac{1}{2} \left[ \frac{\Gamma (j + 4) \Gamma (k)}{
\Gamma (j) \Gamma (k + 4)} - 1 \right],
\end{eqnarray}
we get the quantity in question
\begin{eqnarray}
{^{GG}\!w_{jk}}
\!\!\!&=&\!\!\! - 2C_A \sum_{i = 1}^{3} {^{GG}\!w^{(i)}_{jk}}
= - 2 C_A [ 1 + ( - 1)^{j - k} ] \theta_{j - 2,k}
( 3 + 2k ) \\
&\times&\!\!\!\left\{
2 A_{jk} + ( A_{jk} - \psi (j+2) + \psi(1) )
\left[
\frac{\Gamma (j + 4)\Gamma (k)}{\Gamma (j)\Gamma (k + 4)} - 1
\right]
+ 2 (j - k)( j + k + 3 )
\frac{\Gamma (k)}{\Gamma (k + 4)}
\right\} , \nonumber
\end{eqnarray}
where we have introduced the matrix $\hat A$ the elements of which
are defined by
\begin{equation}
A_{jk} = \psi\left( \frac{j + k + 4}{2} \right)
- \psi\left( \frac{j - k}{2} \right)
+ 2 \psi ( j - k ) - \psi ( j + 2 ) - \psi(1) .
\end{equation}

The matrices for the $QQ$ and $GQ$ cases can be found analogously and were
derived in our previous studies \cite{Mue94,BelMue98}:
\begin{eqnarray}
{^{QQ}\!w}_{jk}
&=&
- 2 C_F \left[ 1 + (-1)^{j-k} \right] \theta_{j-2,k}
( 3 + 2 k ) \nonumber\\
&\times&\left\{
2 A_{jk} + ( A_{jk} - \psi( j + 2 ) + \psi(1) )
\frac{(j - k)(j + k + 3)}{( k + 1 )( k + 2 )}
\right\} , \\
\label{w-GQ}
{^{GQ}\!w}_{jk}
&=&
- 2 C_F \left[ 1 + (-1)^{j-k} \right] \theta_{j-2,k} ( 3 + 2k )
\frac{1}{6}
\frac{(j - k)(j + k + 3)}{( k + 1 ) ( k + 2 )} .
\end{eqnarray}
In addition ${^{QG}\!w} = 0$.

\section{Two-loop anomalous dimensions and evolution kernels.}

From the consistency relation (\ref{const-1}) we can now derive
the equation which defines the off-diagonal elements of the two-loop
anomalous dimension matrix via the one-loop quantities evaluated in
the preceding sections. Namely, we get ($j > k$)
\begin{equation}
a (j,k) \hat\gamma_{jk}^{{\rm ND}(1)}
= \hat\gamma_{j}^{(0)} \hat\gamma_{jk}^{c(0)}
- \hat\gamma_{jk}^{c(0)} \hat\gamma_{k}^{(0)}
+ \beta_0 \left(
\hat\gamma_{j}^{(0)} \hat P_Q - \hat P_Q \hat\gamma_{k}^{(0)}
\right) b_{jk} .
\end{equation}
The complete entry which governs the evolution in NLO is
defined by the sum $\hat\gamma_{jk}^{(1)} = \hat\gamma_j^{(1)}
\delta_{jk} + \hat\gamma_{jk}^{{\rm ND}(1)}$, with $\hat\gamma_j^{(1)}$
being the two-loop forward anomalous dimensions \cite{Fur80,Nee96,Vog97}.
Defining new matrices $d_{jk} = b_{jk}/a(j,k)$ and $g_{jk}
= w_{jk}/a(j,k)$, we can rewrite the above equality in components
\begin{eqnarray}
\label{andimND-QQ}
{^{QQ}\!\gamma}_{jk}^{{\rm ND}(1)}
&=&
\left(
{^{QQ}\!\gamma}_{j}^{(0)} - {^{QQ}\!\gamma}_{k}^{(0)}
\right)
\left\{
d_{jk}
\left(
\beta_0 - {^{QQ}\!\gamma}_{k}^{(0)}
\right)
+ {^{QQ}\!g}_{jk}
\right\} \\
&&\qquad\qquad\qquad\qquad\quad -
\left(
{^{QG}\!\gamma}_{j}^{(0)} - {^{QG}\!\gamma}_{k}^{(0)}
\right) d_{jk}
{^{GQ}\!\gamma}_{k}^{(0)}
+ {^{QG}\!\gamma}_{j}^{(0)} {^{GQ}\!g}_{jk}, \nonumber\\
\label{andimND-QG}
{^{QG}\!\gamma}_{jk}^{{\rm ND}(1)}
&=&
\left(
{^{QG}\!\gamma}_{j}^{(0)} - {^{QG}\!\gamma}_{k}^{(0)}
\right)
d_{jk}
\left( \beta_0 - {^{GG}\!\gamma}_{k}^{(0)} \right)
- \left(
{^{QQ}\!\gamma}_{j}^{(0)} - {^{QQ}\!\gamma}_{k}^{(0)}
\right)
d_{jk} {^{QG}\!\gamma}_{k}^{(0)} \\
&&\qquad\qquad\qquad\qquad\quad +
{^{QG}\!\gamma}_{k}^{(0)} {^{GG}\!g}_{jk}
-
{^{QQ}\!g}_{jk} {^{QG}\!\gamma}_{k}^{(0)} , \nonumber\\
\label{andimND-GQ}
{^{GQ}\!\gamma}_{jk}^{{\rm ND}(1)}
&=&
\left(
{^{GQ}\!\gamma}_{j}^{(0)} - {^{GQ}\!\gamma}_{k}^{(0)}
\right) d_{jk}
\left(
\beta_0 - {^{QQ}\!\gamma}_{k}^{(0)}
\right)
-
\left(
{^{GG}\!\gamma}_{j}^{(0)} - {^{GG}\!\gamma}_{k}^{(0)}
\right) d_{jk}
{^{GQ}\!\gamma}_{k}^{(0)} \\
&&\qquad\qquad\qquad\qquad\quad +
{^{GQ}\!\gamma}_{j}^{(0)} {^{QQ}\!g}_{jk}
-
{^{GG}\!g}_{jk} {^{GQ}\!\gamma}_{k}^{(0)}
+
\left(
{^{GG}\!\gamma}_{j}^{(0)} - {^{QQ}\!\gamma}_{k}^{(0)}
\right)
{^{GQ}\!g}_{jk} , \nonumber\\
\label{andimND-GG}
{^{GG}\!\gamma}_{jk}^{{\rm ND}(1)}
&=&
\left(
{^{GG}\!\gamma}_{j}^{(0)} - {^{GG}\!\gamma}_{k}^{(0)}
\right)
\left\{
d_{jk}
\left(
\beta_0 - {^{GG}\!\gamma}_{k}^{(0)}
\right)
+ {^{GG}\!g}_{jk}
\right\} \\
&&\qquad\qquad\qquad\qquad\quad -
\left(
{^{GQ}\!\gamma}_{j}^{(0)} - {^{GQ}\!\gamma}_{k}^{(0)}
\right) d_{jk}
{^{QG}\!\gamma}_{k}^{(0)}
-
{^{GQ}\!g}_{jk}{^{QG}\!\gamma}_{k}^{(0)}. \nonumber
\end{eqnarray}
Here the leading order anomalous dimensions of conformal operators
read \cite{Cha80,ShiVys81,Ohr81,BelMue98}
\begin{eqnarray}
\label{anomalous-dimensions}
{^{QQ}\!\gamma}_{j}^{(0)}
&=&
- C_F \left(
3 + \frac{2}{( j + 1 )( j + 2 )} - 4 \psi( j + 2 ) + 4 \psi(1)
\right)
\\
{^{QG}\!\gamma}_{j}^{(0)}
&=&
\frac{-24 N_f T_F}{j( j + 1 )( j + 2 )( j + 3 )}
\times \left\{
j^2 + 3 j + 4,\quad \mbox{for even parity}
\atop
j( j + 3 ),\quad\quad  \mbox{for odd parity}
\right.
\\
{^{GQ}\!\gamma}_{j}^{(0)}
&=&
\frac{-C_F}{3( j + 1 )( j + 2 )}
\times \left\{
j^2 + 3 j + 4 ,\quad \mbox{for even parity}
\atop
j ( j + 3 ),\quad\quad \mbox{for odd parity}
\right.
\\
\label{last-AD}
{^{GG}\!\gamma}_{j}^{(0)}
&=&
- C_A \left(
- 4 \psi( j + 2 ) + 4 \psi(1) - \frac{\beta_0}{C_A}
\right) \\
&&\hspace{3cm}-
\frac{8 C_A}{j( j + 1 )( j + 2 )( j + 3 )}
\times \left\{
j^2 + 3 j + 3,\quad \mbox{for even parity}
\atop
j( j + 3 ),\quad\quad  \mbox{for odd parity}
\right. . \nonumber
\end{eqnarray}
and for the transversity sector \cite{BelMue97a,JiHoo98}
\begin{equation}
{^{QQ}\!\gamma}_{j}^{(0)}
=
- C_F \left( 3 - 4 \psi( j + 2 ) + 4 \psi(1) \right) .
\end{equation}
Note that Eqs.\ (\ref{anomalous-dimensions}-\ref{last-AD}) fulfill the
Dokshitzer SUSY relation \cite{Dok77,BFKL85} (here $C_A = C_F = 2 T_F$)
\begin{equation}
{^{QQ}\!\gamma}_{j}^{(0)}
+
\frac{6}{j}{^{GQ}\!\gamma}_{j}^{(0)}
=
\frac{j}{6}{^{QG}\!\gamma}_{j}^{(0)}
+
{^{GG}\!\gamma}_{j}^{(0)}.
\end{equation}
The prefactors $j/6\ (6/j)$ result from the conventional normalization
of the Gegenbauer polynomials.

When translated into the language of the evolution kernels their
off-diagonal parts in NLO read:
\begin{eqnarray}
{^{QQ} V}^{{\rm ND}(1)}
\!\!\!&=&\!\!\! - ( \I - \D )
\biggl\{
{^{QQ}\dot{V}} \otimes \left( {^{QQ} V}^{(0)} + \frac{\beta_0}{2} I \right)
\\
&&\!\!\!+ {^{QQ}\!g} \otimes {^{QQ} V}^{(0)}
- {^{QQ} V}^{(0)}  \otimes {^{QQ}\!g}
+ {^{QG}\dot{V}} \otimes {^{GQ}V}^{(0)}
- {^{QG} V}^{(0)} \otimes {^{GQ}\!g}
\biggr\} , \nonumber\\
{^{QG}V}^{{\rm ND}(1)}
\!\!\!&=&\!\!\! - ( \I - \D )
\biggl\{
{^{QG}\dot{V}} \otimes \left( {^{GG} V}^{(0)} + \frac{\beta_0}{2} I \right)
\\
&&\!\!\!+ {^{QQ}\dot{V}} \otimes {^{QG} V}^{(0)}
+ {^{QQ}\!g} \otimes {^{QG}V}^{(0)}
- {^{QG}V}^{(0)} \otimes {^{GG}\!g}
\biggr\} , \nonumber\\
{^{GQ} V}^{{\rm ND}(1)}
\!\!\!&=&\!\!\! - ( \I - \D )
\biggl\{
{^{GQ}\dot{V}} \otimes \left( {^{QQ} V}^{(0)} + \frac{\beta_0}{2} I \right)
\\
&&\!\!\!+ {^{GG}\dot{V}} \otimes {^{GQ} V}^{(0)}
- {^{GQ}V}^{(0)} \otimes {^{QQ}\!g}
+ {^{GG}\!g} \otimes {^{GQ}V}^{(0)}
- {^{GG}V}^{(0)} \otimes {^{GQ}\!g}
+ {^{GQ}\!g} \otimes {^{QQ}V}^{(0)}
\biggr\} , \nonumber\\
\label{two-loop-GGV}
{^{GG} V}^{{\rm ND}(1)}
\!\!\!&=&\!\!\! - ( \I - \D )
\biggr\{
{^{GG}\dot{V}} \otimes \left( {^{GG} V}^{(0)} + \frac{\beta_0}{2} I \right)
\\
&&\!\!\!+ {^{GG}\!g} \otimes {^{GG}V}^{(0)}
- {^{GG}V}^{(0)} \otimes {^{GG}\!g}
+ {^{GQ}\dot{V}} \otimes {^{QG}V}^{(0)}
+ {^{GQ}\!g} \otimes {^{QG}V}^{(0)}
\biggr\} . \nonumber
\end{eqnarray}
Here we use a shorthand $\otimes = \int_{0}^{1} dz$ so that for any
two test functions $\tau_1 \otimes \tau_2 = \int_{0}^{1} dz
\tau_1 (x, z) \tau_2 (z, y)$. As usual, the $( \I - \D )$-projector
extracts the off-diagonal matrix elements of any function in the
basis of the Gegenbauer polynomials \cite{BelMue98}.

Let us mention that the $\gamma_5$-problem which might arise
in dimensional regularization does not show up in the off-diagonal
elements in NLO. This follows from the fact that the finite
renormalization constant which restore the chiral invariance
broken in the t' Hooft-Veltman-Breitenlohner-Maison scheme is
diagonal in the basis of the Gegenbauer polynomials and thus
cannot affect the results given above \cite{MueTer97}.

The normalization of the Gegenbauer moments is defined as follows
\begin{equation}
\int_{0}^{1} dx
C^{\nu (A)}_j (2x - 1)
{^{AB}\!K} ( x, y )
=
\sum_{k = 0}^{j} {^{AB}\!K}_{jk}
C^{\nu (B)}_k
(2y - 1),
\end{equation}
so that we have the correspondence $K = V$ and $K_{jk} =
- \frac{1}{2} \gamma_{jk}$, $K = (\I - \D) g$ and $K_{jk}
= \theta_{j - 2,k} g_{jk}$, $K = (\I - \D) \dot V$ and $K_{jk} =
\theta_{j - 2,k} ({^{AB}\!\gamma}_j^{(0)}-{^{AB}\!\gamma}_k^{(0)}) d_{jk}$.
The leading order ER-BL kernels have the following form
\begin{equation}
{^{AB} V}^{(0)}( x, y )
=
\theta( y - x )
\, {^{AB}\!F} ( x, y ) \pm
\left( x \to \bar x \atop  y \to \bar y \right) ,
\mbox{\ for\ }
\left\{ {\mbox{$A=B$} \atop \mbox{$A \not= B$}}\right. ,
\end{equation}
with\footnote{Note that the terms with $\delta$-function are understood
in the following way $\delta(x - y)\left[ \theta (y - x) + \theta (x - y)
\right] = \delta(x - y)$.} the following kernels for the parity-even
\cite{BroLep80,EfrRad80,Cha80,Rad96,Ji96,Blum97,BelMue98} and -odd
\cite{Ohr81,BaiGro81,Rad96,Ji96,Blum97,BelMue98} channels
\begin{eqnarray}
{^{QQ}\!F} ( x, y )
&=& C_F \frac{x}{y}
\left[ 1 + \frac{1}{(y-x)_+} + \frac{3}{2} \delta(x-y) \right] ,
\\
\label{V-QG}
{^{QG}\!F} ( x, y )
&=&
2 N_f T_F \frac{x}{y^2 \bar y} \left\{
2 x - y - 1 , \quad \mbox{\ for even parity} \atop
- \bar y , \qquad\qquad \mbox{\ for odd parity} \right. ,
\\
\label{V-GQ}
{^{GQ}\!F} ( x, y )
&=& C_F \frac{x}{y} \left\{
2 y - x , \quad \mbox{\ for even parity} \atop
x , \qquad\quad \mbox{\ for odd parity} \right. ,
\\
{^{GG}\!F} ( x, y )
&=& C_A
\frac{x^2}{y^2}
\left[
\frac{1}{(y-x)_+} - \frac{1}{2}\frac{\beta_0}{C_A} \delta (x - y)
\right]
+ 2 C_A \frac{x^2}{y^2}
\left\{
\bar x + y ( 1 + 2 \bar x ) , \quad \mbox{\ for even parity} \atop
1 , \qquad\qquad\qquad\ \mbox{\ for odd parity} \right. .
\nonumber\\
\end{eqnarray}
While for the chiral odd sector we have \cite{BelMue97a} (see also
\cite{JiHoo98})
\begin{equation}
{^{QQ}\!F} ( x, y )
= C_F \frac{x}{y} \left[ \frac{1}{(y-x)_+}
- \frac{1}{2} \delta (x - y) \right] .
\end{equation}
For the dotted kernels we get \cite{BelMue98}
\begin{eqnarray}
\label{dot-ker}
{^{AB} \dot{V}} ( x, y )
&=& \theta(y-x) \,
\left[
{^{AB}\!F}(x,y) \ln\frac{x}{y}
+
\Delta {^{AB}\!\dot F} ( x, y )
\right]
\pm
\left( x \to \bar x \atop y \to \bar y \right),
\mbox{\ for\ } \left\{ {\mbox{$A=B$} \atop \mbox{$A\not=B$}} \right. ,
\end{eqnarray}
with
\begin{eqnarray}
\label{AddKer}
&&\Delta {^{GQ}\!{\dot F}^V}
= 2 C_F
\left[
x \ln y - \bar x \ln \bar x
\right] ,
\quad
\Delta {^{QG}\!{\dot F}^V}
= 4 T_F N_f \frac{1}{y \bar y}
\left[
x \ln y - \bar x \ln \bar x
\right] , \\
&&\qquad\qquad\qquad\qquad\Delta {^{GG}\!{\dot F}^V}
= 2 C_A \frac{x^2}{y^2}(y - x) .
\end{eqnarray}
In all other cases $\Delta {^{AB}\!\dot F} ( x, y ) = 0$.


\begin{figure}[t]
\begin{center}
\vspace{5.0cm}
\hspace{-1.5cm}
\mbox{
\begin{picture}(0,220)(270,0)
\put(0,-30)                    {
\epsffile{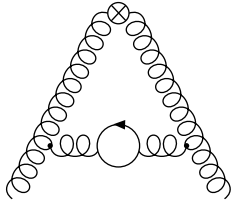}
                               }
\end{picture}
}
\end{center}
\vspace{-10cm}
\caption{\label{GG-bubble-ins} This diagram generates the $N_f$-dependent
part of the term $\beta_0 {^{GG} \dot{V}} ( x, y )$ in the two-loop
ER-BL kernel.}
\end{figure}

A consistency check for the $\beta_0$-dependent part of the mixed
$GQ$ and $QG$ channels was given in Ref.\ \cite{BelMue98} by an explicit
calculation of the simplest two-loop diagrams with fermion bubble
insertion. Here we mention that a similar analysis can be easily done
for the pure gluonic sector. First we should mention that
there is no net $N_f$-dependence of the off-diagonal kernel
${^{GG} V}^{{\rm ND}(1)}(x, y)$. However, one can argue that
among other graphs the diagram in Fig.\ \ref{GG-bubble-ins} produces
the $N_f$-part of the complete QCD $\beta$-function in
Eq.\ (\ref{two-loop-GGV}). Since we know that
the dotted kernel for the parity odd sector, which arises from the
logarithmic modification of the leading order kernel, is correct,
we can just consider the difference between the vector and
axial vector channels. Then, neglecting in the expression
for diagram in Fig.\ \ref{GG-bubble-ins} the piece which comes,
after subtraction of subdivergencies from the ${\cal O}(\epsilon)$-effects,
(the latter generate only diagonal contributions to the kernel), we
get the following result for the light-cone position dotted kernel:
\begin{equation}
\label{GG-bubble}
{^{GG}\!\dot\K^{V-A}} (y, z) = 12 C_A y z \ln (1 - y - z).
\end{equation}
This result is quite obvious, since the only effect of the fermion
bubble insertion into the gluon line is the shift of the power of the
denominator by the amount $\epsilon$. Due to the fact that the weight
factor associated with this line is $(1 - y - z)^{- \epsilon}$,
the first non-trivial term in the Taylor expansion gives us the
result (\ref{GG-bubble}), where the factor $12C_A yz$ is just the
difference between the leading order gluonic kernels
(\ref{GGV-kernel},\ref{GGA-kernel}). By Fourier transformation
to the momentum space we easily obtain
\begin{equation}
{^{GG} \dot V^{V - A}} (x, y)
= \theta (y - x)
\left[
\Delta {^{GG}\!{\dot F}^V} (x, y)
+ \left( {^{GG}\! F^V} (x, y) - {^{GG}\! F^A} (x, y) \right)
\left( \ln\frac{x}{y} - \frac{11}{6} \right)
\right]
+ \left( { x \to \bar x \atop y \to \bar y } \right).
\end{equation}
Getting rid of the remaining diagonal piece, we obtain the result displayed
in Eqs.\ (\ref{dot-ker},\ref{AddKer}). It is a trivial task to check that
the moments of this kernel is equal to $( {^{GG}\!\gamma}_j^{(0)} -
{^{GG}\!\gamma}_k^{(0)})d_{jk}$.

The last but not the least ingredients for the off-diagonal part of
the two-loop evolution kernels are the $g$-functions, which are given
by the formulae
\begin{eqnarray}
\label{set-g-kernels}
{^{QQ}\!g}(x,y)
&=& - C_F
\left[ \theta( y - x )
\frac{ \ln \left( 1 - \frac{x}{y} \right) }{y - x} +
\left( x \to \bar{x} \atop y\to \bar{y} \right) \right]_+,
\qquad
{^{QG}\!g}(x,y) = 0 ,
\\
{^{GQ}\!g}(x,y)
&=& - C_F
\left[
\theta( y - x )
\ln \left( 1 - \frac{x}{y} \right) -
\left( x \to \bar{x} \atop y\to \bar{y} \right)
\right] ,
\qquad
{^{GG}\!g}(x,y)
= \frac{C_A}{C_F} {^{QQ}\!g}(x,y) . \nonumber
\end{eqnarray}
These results were obtained by solving the second order differential
equations for the $QQ$ and $GG$ channels and by explicit summation
of the infinite series of the Gegenbauer polynomials for the $GQ$ one.

\section{Solution of the two-loop evolution equations.}

So far we have determined the matrix elements of the anomalous
dimension matrix of the conformal operators in NLO. In order
to solve the renormalization group equation one should determine
its eigenvectors (the eigenvalues are given by the diagonal elements
and coincide with known forward anomalous dimensions). This section
is devoted to the solution of this problem. It will be shown here
that they are determined again by the special conformal anomaly
matrix introduced previously.

Beyond leading order one has to perform an additional shuffling
of operators in order to get the eigenstates of the renormalization
group equation. Therefore, let us consider the set of operators
$\widetilde{\cal O}_{jl}$
\begin{eqnarray}
\label{rotatedCO}
[{\cal O}_{jl}]
= \sum_{k=0}^{j} \hat B_{jk} (\g) [\widetilde {\cal O}_{kl}] ,
\end{eqnarray}
where the $\hat B$-matrix defines the rotation to the diagonal basis.
This new set of operators satisfies the renormalization group
equation
\begin{equation}
\mu\frac{d}{d\mu}
[\widetilde {\cal O}_{jl}]
= - \hat\gamma^{\rm D}_j \ [\widetilde {\cal O}_{jl}],
\quad \mbox{and} \quad
\hat\gamma^{\rm D}_j
= \left(
{ {^{QQ}\!\gamma}^{\rm D}_j \ {^{QG}\!\gamma}^{\rm D}_j
\atop
{^{GQ}\!\gamma}^{\rm D}_j \ {^{GG}\!\gamma}^{\rm D}_j }
\right),
\end{equation}
with the following solution
\begin{equation}
\label{RGsolution}
[\widetilde {\cal O}_{jl} (\mu^2)]
= {\cal T} \exp
\left\{
- \frac{1}{2} \int_{\mu_0^2}^{\mu^2} \frac{d\tau}{\tau}
\hat\gamma^{\rm D}_j (\alpha_s (\tau))
\right\}
[\widetilde {\cal O}_{jl} (\mu_0^2)] .
\end{equation}
Since the matrices $\hat\gamma^{\rm D}_j$ do not commute with each
other, we have introduced the ${\cal T}$-ordered exponential.

Substituting Eq.\ (\ref{rotatedCO}) into the conformal spin expansion
\begin{equation}
\phi (x) = \sum_{j = 0}^{\infty} \phi_j (x)
\langle h'|
[{\cal O}_{jj}]
| h \rangle ,
\end{equation}
for the wave function $\phi (x)$ which is defined as a two-dimensional
vector
\begin{equation}
\phi (x)
= \left(
{ {^Q \phi}(x) \atop {^G \phi} (x)}
\right),
\end{equation}
and with the conformal waves being Gegenbauer polynomials
\begin{equation}
\label{LOpartialwaves}
\phi_j (x)
=
\left(
{ w ( x | \frac{3}{2} ) / N_j (\frac{3}{2})
C_j^{\frac{3}{2}} (2x - 1)
\atop
w ( x | \frac{5}{2} ) / N_{j - 1} (\frac{5}{2})
C_{j - 1}^{\frac{5}{2}} (2x - 1) }
\right) ,
\end{equation}
we obtain the expansion in terms of the eigenfunctions
$\phi_j (x | \alpha_s)$ of the ER-BL evolution equation beyond
leading order
\begin{equation}
\phi (x) = \sum_{j = 0}^{\infty} \phi_j (x | \alpha_s)
\langle h'|
[\widetilde {\cal O}_{jj}]
| h \rangle ,
\quad \mbox{with} \quad
\phi_j (x | \alpha_s) = \sum_{k = j}^{\infty} \phi_k (x) \hat B_{kj} (\g).
\end{equation}
As it is seen the eigenfunctions are generalized to non-polynomial
functions and corrections to the leading order partial waves
(\ref{LOpartialwaves}) are determined entirely by the $\hat B$-matrix
which satisfies the following differential equation:
\begin{equation}
\label{B-equation}
\beta (\g) \frac{\partial}{\partial\g} \hat B (\g)
+ [\hat \gamma^{\rm D}, \hat B (\g)]_-
+ \hat \gamma^{\rm ND} \hat B (\g) = 0.
\end{equation}

As will be shown in the following two sections, the formalism we have
described allows to find the corrections to the eigenfunctions
analytically, since the $\hat B$-matrix is defined in terms of the
conformal anomaly $\hat\gamma^c$ and the $\beta$-function. Below we
treat two different cases separately: the hypothetical conformal limit
and the case with running coupling constant.

\subsection{$\beta (\g) = 0$.}

As we have observed the problem of finding the corrections to the
eigenfunctions is reduced to the evaluation of the $\hat B$-matrix
which satisfies the first order differential equation (\ref{B-equation}).
In the conformal limit of the theory it simplifies to
\begin{equation}
[\hat \gamma^{\rm D}, \hat B (\g)]_-
+ \hat \gamma^{\rm ND} \hat B (\g) = 0.
\end{equation}
We can construct a recursive perturbative solution of this equation,
$\hat B = \sum_{\ell = 0} \hat B^{(\ell)}$. Since in leading order
the operators with different conformal spin $j$ do not mix with each
other, it provides us with condition $\hat B^{(0)} = \hat 1$. Combining
these results give us an inhomogeneous equation for $\hat B^{(\ell)}$
with the source $\hat\gamma^{\rm ND} \hat B^{(\ell - 1)}$
treated as perturbation
\begin{equation}
\label{BconformalEq}
[\hat \gamma^{\rm D}, \hat B^{(\ell)} (\g)]_-
= - \hat \gamma^{\rm ND} \hat B^{(\ell - 1)} (\g) ,
\qquad (\ell \geq 1) .
\end{equation}
For the first non-trivial term $\hat B^{(1)}$, which is the only one
we need to the accuracy we are limited for, we have the following
solution
\begin{equation}
\hat B^{(1)}_{jk} = - \frac{1}{a(j, k)} \hat\gamma^{c}_{jk},
\end{equation}
where on the RHS of Eq.\ (\ref{BconformalEq}) we insert the expression
for $\hat\gamma^{\rm ND}$ in terms of the special conformal anomaly
matrix. Thus, the new two-loop eigenfunctions read
\begin{equation}
\phi_j (x | \alpha_s) = \sum_{k = j}^{\infty}
\phi_k (x)
\left\{
\hat 1_{jk}
- \frac{1}{a(k, j)} \hat\gamma^{c}_{kj}
\right\} .
\end{equation}
This result can be conveniently written in the form of a convolution,
so that we have
\begin{equation}
\left(
{{^Q\!\phi_j}
\atop
{^G\!\phi_j}}
\right)
(x | \alpha_s)
= \left\{
\left(
{ 1 \ 0 \atop 0 \ 1}
\right)
\delta (x - y)
+ \frac{\alpha_s}{2\pi}
\left(
{
{^{QQ} {\mit\Phi}} \ {^{QG} {\mit\Phi}}
\atop
{^{GQ} {\mit\Phi}} \ {^{GG} {\mit\Phi}}
}
\right)
(x, y)
\right\}
\otimes
\left(
{{^Q\!\phi_j}
\atop
{^G\!\phi_j}}
\right)
(y) .
\end{equation}
Here the matrix-valued function ${\mit\Phi} (x, y)$ is expressed in
terms of kernels derived in the preceding sections
\begin{equation}
{\mit\Phi} (x, y)
= - ( \I - \D )
\left\{ S (x, z) \otimes V^{(0)} (z, y) + g (x, y) \right\},
\end{equation}
which thus define the complete set of corrections to the eigenfunctions of
NLO evolution equation. Here the $S$-operator generates the shift in the
index of the Gegenbauer polynomials:
\begin{equation}
S(x, y) \otimes [y \bar y]^{\nu - \frac{1}{2}}
C^\nu_j (2y - 1)
= \left. \frac{d}{d\rho} \right|_{\rho = 0}
[x \bar x]^{\nu - \frac{1}{2} + \rho} C^{\nu + \rho}_j (2x - 1).
\end{equation}
We can also recover the dependence on $\beta_0$ which is related
to the shift of the index of Gegenbauer polynomials and not to
the effects of a running $\alpha_s$. From the definition of the
conformal anomaly matrix it follows that this reduces to a mere
substitution
\begin{equation}
V^{(0)} (x, y) \Rightarrow V^{(0)} (x, y)
+ \frac{\beta_0}{2} \delta(x - y) .
\end{equation}
Note that the results for the $QQ$ and $QG$ sectors were used by us
previously for the evaluation of NLO corrections to the amplitudes
of the non-forward processes making use of the conformal covariant
operator product expansion \cite{BelMue97a,BelSch98}.

Let us now address the case with $\beta (\g) \neq 0$.

\subsection{$\beta (\g) \neq 0$.}

In the case of a running coupling constant the solution of the
corresponding differential equation (\ref{B-equation}) is more
complicated. First, we should specify the boundary condition for
$\hat B (\g)$. Note that in the previous case of a fixed coupling we are
naturally led to a scheme with constant $\hat B$. The minimization
of the radiative corrections corresponds to the choice $\hat B (\g_0) = 1$
that means the absence of the former at the reference point $\mu_0$
($\g_0 = \g (\mu_0^2)$). The most important advantage of this requirement
is that contrary to our previous discussion the initial condition for the
solution of the RG equation (\ref{RGsolution}) can be easily determined
by composing ordinary Gegenbauer moments of the initial wave function
since
\begin{equation}
[ {\cal O}_{jl} (\mu_0) ]
= [ \widetilde{\cal O}_{jl} (\mu_0) ] .
\end{equation}
To the required two-loop accuracy we can write the equation for the
recursive functions $\hat B_{jk} (\g) = \hat 1_{jk} +
\hat B^{(1)}_{jk} (\g)$
\begin{equation}
\beta (\g) \frac{\partial}{\partial\g} \hat B^{(1)} (\g)
+ [\hat \gamma^{\rm D}, \hat B^{(1)} (\g)]_-
+ \hat \gamma^{\rm ND} = 0.
\end{equation}
The solution can be written in the form
\begin{equation}
\hat B^{(1)}_{jk} (\g)
= - {\cal G} \int_{\g_0}^{\g} \frac{d \g'}{\beta (\g')}
\exp\left\{
- \int_{\g'}^{\g} \frac{d \g''}{\beta (\g'')}
\hat\gamma^{\rm D}_j (\g'')
\right\}
\hat\gamma^{\rm ND}_{jk} (\g')
\exp\left\{
- \int_{\g}^{\g'} \frac{d \g''}{\beta (\g'')}
\hat\gamma^{\rm D}_k (\g'')
\right\} ,
\end{equation}
where we have introduced ${\cal G}$-ordering along the integration path.

Recapitulating the results derived here we have found the analytical
solution of the two-loop ER-BL evolution equation and thus can study
explicitly the effects of NLO corrections to the wave functions
as well as to the non-forward parton distribution functions.

\section{Conclusion.}

To summarize, in this paper we have generalized the formalism,
developed previously for the evaluation of the two-loop corrections
to the non-diagonal part of the singlet exclusive evolution kernels
in an Abelian theory, to the QCD case. We stress once more that the
diagonal entries are determined entirely by the known DGLAP splitting
functions. An enormous simplification of the calculations we have
presently performed is due to the fact that the ${\cal O}
(\alpha_s^2)$-corrections to the evolution kernels are
obtained from the evaluation of the one-loop renormalization
matrices of the conformal operators. To obtain these we have used
specific Feynman rules which result from the
study of the special conformal Ward identities. The latter control the
way how the conformal symmetry is broken in gauge theories. We would like
to point out that our results can be used as a guide for a simple
convolution-type representation \cite{MikRad86} of the full evolution
kernels calculated in the direct way. Moreover, numerical simulations
previously done in the $QQ$-channel \cite{Kat85} can serve as a
consistency check.

As a by-product of our analysis we gained some insight into the structure
of the counterterms for the product of two renormalized composite
operators, a problem which has not been solved so far. Evidently,
the complete solution of this task deserves further study in order to
put it on the same grounds as Joglekar-Lee renormalization theory of
composite operators \cite{JogLee75}.

We have observed universality of the LO special conformal anomalies
in the sense that they do not depend on parity and chirality but
only on the particles involved. The difference in the off-diagonal
elements of the NLO anomalous dimension matrices arises as a result
of different dilatation anomalies, i.e.\ LO anomalous dimensions
of the conformal operators.

Therefore, at present stage we have all the necessary perturbative input
for explicit studies of evolution effects for the wave function
beyond leading order in the flavour singlet channel. However, one
should mention that convergence of the partial wave expansion
in particular regions of the phase space is numerically difficult to
achieve. A study of the magnitude of the effects due to two-loop
evolution for the non-singlet non-forward distribution functions will
be addressed by us elsewhere \cite{BelMueLeoSch98}.

Moreover, now it is possible to perform a complete NLO analysis for
deeply virtual Compton scattering in the leading twist-2 approximation,
since the one-loop coefficient functions were determined earlier
\cite{JiOs97,BelMue97a,Mank97,BelSch98,JiOs98}. One can also
include the effects of non-leading twist power corrections which
were already estimated by us using renormalon based techniques
\cite{BelSch98}.

\vspace{0.5cm}

We would like to thank A. Sch\"afer for kind support, careful
reading of the manuscript and useful comments. A.B. was supported
by the Alexander von Humboldt Foundation and partially by Russian
Foundation for Fundamental Research, grant N 96-02-17631. He also uses
the opportunity to thank Mr. J. Tran Thanh Van and the Organizing
Committee of XXXIIIrd Rencontres de Moriond: ``QCD and high energy
hadronic interactions'' for a grant in the framework of `Training and
Mobility of Researches' which allowed him to attend such a fruitful
conference.

\appendix

\setcounter{section}{0}
\setcounter{equation}{0}
\renewcommand{\theequation}{\Alph{section}.\arabic{equation}}

\section{Feynman rules for the operator insertions.}
\label{FR}

In this appendix we list the Feynman rules for the operator insertions
which enter into the special conformal Ward identities as well as the
rules for the light-cone string operators. The latter when sandwiched
between appropriate hadronic states define the non-forward distributions
or wave functions. Namely, these relations read ($\zeta \equiv (p - p')_+$)
\begin{eqnarray}
\label{NFPD-Q}
\langle h'|
{^Q\!{\cal O}^{\mit\Gamma} (\kappa_1, \kappa_2)}
| h \rangle
&=& \int dx
\left[
e^{- i \kappa_1 x - i \kappa_2 (\zeta - x)}
\mp
e^{- i \kappa_2 x - i \kappa_1 (\zeta - x)}
\right]
{^Q\!{\cal O}^{\mit\Gamma} (x, \zeta)},\\
\label{NFPD-G}
\langle h'|
{^G\!{\cal O}^{\mit\Gamma} (\kappa_1, \kappa_2)}
| h \rangle
&=& \frac{1}{2} \int dx
\left[
e^{- i \kappa_1 x - i \kappa_2 (\zeta - x)}
\pm
e^{- i \kappa_2 x - i \kappa_1 (\zeta - x)}
\right]
{^G\!{\cal O}^{\mit\Gamma} (x, \zeta)},
\end{eqnarray}
where the non-local operators are defined in Eqs.\
(\ref{treeQLRO},\ref{treeGLRO}). When $\zeta = 1$ and one of the hadronic
states is replaced by the vacuum the ${\cal O}$'s coincide with usual
singlet wave functions ${^Q\!{\cal O}(x, \zeta = 1)} = {^Q\!\phi (x)}$
and ${^G\!{\cal O}(x, \zeta = 1)} = {^G\!\phi (x)}$, which obey the
evolution equation (\ref{ER-BLequation}).


\begin{figure}[t]
\begin{center}
\vspace{5.0cm}
\hspace{0.2cm}
\mbox{
\begin{picture}(0,220)(270,0)
\put(0,-30)                    {
\epsffile{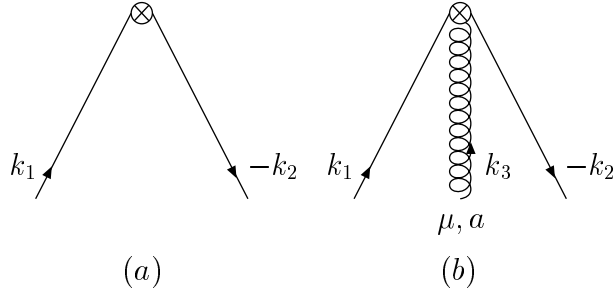}
                               }
\end{picture}
}
\end{center}
\vspace{-9.4cm}
\caption{\label{Feynman-Q} Feynman rules for the quark non-local
string operator.}
\end{figure}


\begin{figure}[t]
\begin{center}
\vspace{4.8cm}
\hspace{-2.5cm}
\mbox{
\begin{picture}(0,75)(270,0)
\put(0,-30)                    {
\epsffile{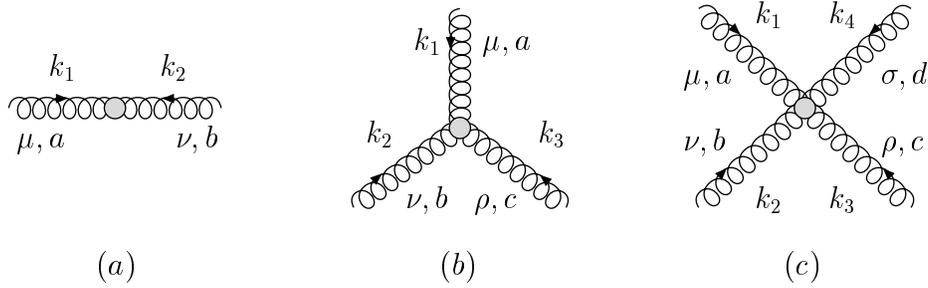}
                               }
\end{picture}
}
\end{center}
\vspace{-4cm}
\caption{\label{Feynman-G} Feynman rules for the gluon operator
insertions. The grey blob stands either for the non-local string
operator ($\otimes$), or the $i[{\cal O}_A^-]$-vertex ($\circ$),
or the usual vertices from the QCD Lagrangian ($\bullet$). }
\end{figure}

For every gluon and quark line we associate the propagators
\begin{equation}
(-i)D^{ab}_{\mu\nu}(k)
=\frac{-i \delta^{ab}}{k^2 + i0}
\left(
g_{\mu\nu} - (1 - \xi)\frac{k_{\mu}k_{\nu}}{k^2}
\right),
\qquad\mbox{and}\qquad
i S (k) = \frac{i \not\! k}{k^2 + i0} ,
\end{equation}
respectively. The Feynman rules for the vertices are given in the next
subsections. The resulting expression, composed out of propagators and
vertices should be integrated with respect to the momentum of every
internal line, i.e.\ multiplied by the factor
\begin{equation}
\int \prod_\ell \frac{d^dk_\ell}{(2\pi)^{\ell\cdot d}}.
\end{equation}

\subsection{Feynman rules for ${^Q\!{\cal O}}(\kappa_1,\kappa_2)$.}

For the quark string operator we have the following rules
\begin{eqnarray}
{{\cal O}} &=& {\mit\Gamma}
\left[
e^{ - i \kappa_1 k_{1+} - i \kappa_2 k_{2+} }
\mp
e^{ - i \kappa_1 k_{2+} - i \kappa_2 k_{1+} }
\right] , \\
{_\Phi{\cal O}}^a_\mu &=& -\g t^a n_\mu {\mit\Gamma}
\left[
e^{ - i \kappa_1 k_{1+} - i \kappa_2 k_{2+} }
\pm
e^{ - i \kappa_1 k_{2+} - i \kappa_2 k_{1+} }
\right]
\frac{ e^{ - i \kappa_2 k_{3+}} - e^{ - i \kappa_1 k_{3+} } }{k_{3+}},
\end{eqnarray}
which are given diagrammatically in Fig.\ \ref{Feynman-Q}. Here the
upper sign in the square brackets stands for ${\mit\Gamma} = \gamma_+,
\sigma_{+\perp} \gamma_5$ and the lower sign for $\gamma_+ \gamma_5$.

\subsection{Feynman rules for ${^G\!{\cal O}}(\kappa_1,\kappa_2)$.}

The gluon non-local operator can be decomposed as
${^G\!{\cal O}} (\kappa_1, \kappa_2) =
{^G\!{\cal O}}_{A} (\kappa_1, \kappa_2)
+ {^G\!{\cal O}}_{NA} (\kappa_1, \kappa_2)
+ {^G\!{\cal O}}_{\Phi} (\kappa_1, \kappa_2)$,
where the subscripts designate the origin of the corresponding
contributions, namely, ($N$)$A$: (non-)Abelian part of the field
strength tensor; $\Phi$: path ordered exponential. In one-loop
approximation the only Feynman rules we need are the following
(crossed circles instead of grey blobs in the diagrams of
Fig.\ \ref{Feynman-G})
\begin{eqnarray}
\label{Abel-gluon}
{_A{\cal O}}_{\mu\nu}^{ab} \!\!\!&=&\!\!\!
\delta^{ab}
\left\{\!\!\!
\begin{array}{c}
g_{\alpha\beta} \\
i \epsilon_{\alpha\beta -+}
\end{array}
\!\!\!\right\}
f_{+ \beta; \mu} (k_1) f_{+ \alpha; \nu} (k_2)
\left[
e^{ - i \kappa_1 k_{1+} - i \kappa_2 k_{2+} }
\pm
e^{ - i \kappa_1 k_{2+} - i \kappa_2 k_{1+} }
\right] , \\
&& \nonumber\\
{_{NA}{\cal O}}_{\mu\nu\rho}^{abc} \!\!\!&=&\!\!\!
i \g f^{abc}
\left\{\!\!\!
\begin{array}{c}
g_{\alpha\beta} \\
i \epsilon_{\alpha\beta -+}
\end{array}
\!\!\!\right\} \\
&&\qquad\times\left\{
n_\mu
\left[
g_{\beta\nu}
f_{+ \alpha; \rho} (k_3) e^{ - i \kappa_1 k_{1+} }
-
g_{\alpha\rho}
f_{+ \beta; \nu} (k_2) e^{ - i \kappa_2 k_{1+} }
\right]
e^{ - i \kappa_1 k_{2+} - i \kappa_2 k_{3+} } \right.
\nonumber\\
&&\qquad+ \,
n_\nu
\left[
g_{\beta\rho}
f_{+ \alpha; \mu} (k_1) e^{ - i \kappa_1 k_{2+} }
-
g_{\alpha\mu}
f_{+ \beta; \rho} (k_3) e^{ - i \kappa_2 k_{2+} }
\right]
e^{ - i \kappa_1 k_{3+} - i \kappa_2 k_{1+} }
\nonumber\\
&&\qquad+ \,
\left. n_\rho
\left[
g_{\beta\mu}
f_{+ \alpha; \nu} (k_2) e^{ - i \kappa_1 k_{3+} }
-
g_{\alpha\nu}
f_{+ \beta; \mu} (k_1) e^{ - i \kappa_2 k_{3+} }
\right]
e^{ - i \kappa_1 k_{1+} - i \kappa_2 k_{2+} }
\pm (\kappa_1 \leftrightarrow \kappa_2)
\right\} , \nonumber\\
&& \nonumber\\
{_{\Phi}{\cal O}}_{\mu\nu\rho}^{abc} \!\!\!&=&\!\!\!
i \g f^{abc}
\left\{\!\!\!
\begin{array}{c}
g_{\alpha\beta} \\
i \epsilon_{\alpha\beta -+}
\end{array}
\!\!\!\right\} \\
&&\qquad\times\left\{
n_\mu f_{+ \alpha; \nu} (k_2) f_{+ \beta; \rho} (k_3)
\left[
e^{ - i \kappa_1 k_{3+} - i \kappa_2 k_{2+} }
\mp
e^{ - i \kappa_1 k_{2+} - i \kappa_2 k_{3+} }
\right]
\frac{ e^{ - i \kappa_2 k_{1+}} - e^{ - i \kappa_1 k_{1+} } }{k_{1+}}
\right.
\nonumber\\
&&\qquad+ \,
n_\nu
f_{+ \alpha; \rho} (k_3) f_{+ \beta; \mu} (k_1)
\left[
e^{ - i \kappa_1 k_{1+} - i \kappa_2 k_{3+} }
\mp
e^{ - i \kappa_1 k_{3+} - i \kappa_2 k_{1+} }
\right]
\frac{ e^{ - i \kappa_2 k_{2+}} - e^{ - i \kappa_1 k_{2+} } }{k_{2+}}
\nonumber\\
&&\qquad+ \,
\left. n_\rho
f_{+ \alpha; \mu} (k_1) f_{+ \beta; \nu} (k_2)
\left[
e^{ - i \kappa_1 k_{2+} - i \kappa_2 k_{1+} }
\mp
e^{ - i \kappa_1 k_{1+} - i \kappa_2 k_{2+} }
\right]
\frac{ e^{ - i \kappa_2 k_{3+}} - e^{ - i \kappa_1 k_{3+} } }{k_{3+}}
\right\}. \nonumber
\end{eqnarray}
Here and below $f_{\alpha \beta; \mu} (k) \varepsilon_\mu = k_\alpha
\varepsilon_\beta - k_\beta \varepsilon_\alpha$ is the Abelian part
of the gluon field strength tensor.

\subsection{Feynman rules for $i[{\cal O}_A^-]$.}

For the operator insertion $i [{\cal O}_A^-]$ which appears in the
special conformal Ward identity we have (empty blobs in the Feynman
rules)
\begin{eqnarray}
&&\hspace{-1.5cm}{_2{\widetilde\V^{ab}_{\mu\nu}}}
= - i \delta^{ab}
f_{\alpha\beta;\mu}(k_1) f_{\alpha\beta;\nu}(k_2)
(2\pi)^4 \, 2 \, i \partial_- \delta^{(4)}
\left( k_1 + k_2 \right) \!, \\
&& \nonumber\\
&&\hspace{-1.5cm}{_3{\widetilde\V^{abc}_{\mu\nu\rho}}}
= - 2 \g f^{abc}
\left\{
(k_1 - k_2)_\rho g_{\mu\nu}
+ (k_2 - k_3)_\mu g_{\nu\rho}
+ (k_3 - k_1)_\nu g_{\rho\mu}
\right\}
(2\pi)^4 \, 2 \, i \partial_- \delta^{(4)}
\left( k_1 + k_2 + k_3 \right) \!, \\
&& \nonumber\\
&&\hspace{-1.5cm}{_4{\widetilde\V^{abcd}_{\mu\nu\rho\sigma}}}
= 2 i \g^2
\bigl\{
f^{abe}f^{cde}
\left(
g_{\mu\rho}g_{\nu\sigma} - g_{\mu\sigma}g_{\nu\rho}
\right)
+ f^{ace}f^{bde}
\left(
g_{\mu\nu}g_{\rho\sigma} - g_{\mu\sigma}g_{\nu\rho}
\right) \\
&&\hspace{4cm}+ f^{ade}f^{cbe}
\left(
g_{\mu\rho}g_{\nu\sigma} - g_{\mu\nu}g_{\rho\sigma}
\right)
\bigr\}
(2\pi)^4 2 \, i \partial_- \delta^{(4)}
\left( k_1 + k_2 + k_3 + k_4 \right) \nonumber.
\end{eqnarray}
It is instructive to compare them with the usual Feynman rules for the
triple and quartic interaction vertices (full points in the graphs):
\begin{eqnarray}
&&\hspace{-1.5cm}{_3{\V^{abc}_{\mu\nu\rho}}}
= \g f^{abc}
\left\{
(k_1 - k_2)_\rho g_{\mu\nu}
+ (k_2 - k_3)_\mu g_{\nu\rho}
+ (k_3 - k_1)_\nu g_{\rho\mu}
\right\}
(2\pi)^4 \, \delta^{(4)} \left( k_1 + k_2 + k_3 \right) \!, \\
&& \nonumber\\
&&\hspace{-1.5cm}{_4{\V^{abcd}_{\mu\nu\rho\sigma}}}
= - i \g^2
\bigl\{
f^{abe}f^{cde}
\left(
g_{\mu\rho}g_{\nu\sigma} - g_{\mu\sigma}g_{\nu\rho}
\right)
+ f^{ace}f^{bde}
\left(
g_{\mu\nu}g_{\rho\sigma} - g_{\mu\sigma}g_{\nu\rho}
\right) \\
&&\hspace{4cm}+ f^{ade}f^{cbe}
\left(
g_{\mu\rho}g_{\nu\sigma} - g_{\mu\nu}g_{\rho\sigma}
\right)
\bigr\}
(2\pi)^4 \delta^{(4)} \left( k_1 + k_2 + k_3 + k_4 \right) \nonumber.
\end{eqnarray}

\subsection{Feynman rules for $\int d^d x \, x_- B_\mu^a (x)
\frac{\delta}{\delta B_\mu^a (x) } [ {^G\!{\cal O}_{jl}} ] $.}

Finally, in the considered approximation we need only the part of
$\int d^d x \, x_- B_\mu^a (x) \frac{\delta}{\delta B_\mu^a (x) }
[ {^G\!{\cal O}} ]$ which survives for $\g = 0$
\begin{eqnarray}
&&\hspace{-0.5cm}\int d^d x \, x_- B_\mu^a (x)
\frac{\delta}{\delta B_\mu^a (x) }
\left. [ {^G\!{\cal O}} (\kappa_1,\kappa_2) ] \right|_{\g = 0 }
=
( \kappa_1 + \kappa_2 )
F^a_{+\mu} (\kappa_2 n)
\left\{\!\!\!
\begin{array}{c}
g_{\mu\nu} \\
i \epsilon_{\mu\nu -+}
\end{array}
\!\!\!\right\}
F^a_{\nu +} (\kappa_1 n) \\
&&+
( B^a_\mu(\kappa_2 n) - n^*_\mu B^a_+(\kappa_2 n) )
\left\{\!\!\!
\begin{array}{c}
g_{\mu\nu} \\
i \epsilon_{\mu\nu -+}
\end{array}
\!\!\!\right\}
F^a_{\nu +} (\kappa_1 n)
-
F^a_{+\mu} (\kappa_2 n)
\left\{\!\!\!
\begin{array}{c}
g_{\mu\nu} \\
i \epsilon_{\mu\nu -+}
\end{array}
\!\!\!\right\}
( B^a_\nu(\kappa_1 n) - n^*_\nu B^a_+(\kappa_1 n) ).
\nonumber
\end{eqnarray}
Thus the Feynman rules for this operator are given by $( \kappa_1 +
\kappa_2 ) {_A{\cal O}}_{\mu\nu}^{ab}(k_1,k_2)$ and the following vertex
for the second and third term
\begin{equation}
i \delta^{ab} \left\{\!\!\!
\begin{array}{c}
g_{\alpha\beta}
\\
i \epsilon_{\alpha\beta - +}
\end{array}
\!\!\!\right\}
\left\{
( g_{\beta \mu} - n_{\beta}^* n_{\mu} ) f_{+ \alpha; \nu} (k_2)
+
( g_{\alpha \nu} - n_{\alpha}^* n_{\nu} ) f_{+ \beta; \mu} (k_1)
\right\}
\left[
e^{ - i \kappa_1 k_{1+} - i \kappa_2 k_{2+} }
\pm
e^{ - i \kappa_1 k_{2+} - i \kappa_2 k_{1+} }
\right] .
\end{equation}

\setcounter{equation}{0}

\section{Renormalization of string operators.}
\label{renormalization}

In this appendix we intend to review the approach for construction
of evolution equations for non-local light-cone operators
developed in Ref.\ \cite{GeyRob84}. Although a lot of results
have been obtained so far using the latter, we did not find in the
literature any comprehensive review which allows a pedestrian to make
these calculations, especially, in the case of non-forward kinematics.
To our opinion this machinery is more general than the background
field formalism \cite{BalBra89}, since the former is not tied to any
particular gauge and thus can be used for studying the form of
gauge-variant counterterms. Next we present the details of the
calculations which were cited in the main text\footnote{In fact, we
have used two different formalisms for our calculations: the one
mentioned above and another technique developed in Ref.\
\protect\cite{Mue97b}. Computer realizations were also implemented
with two different languages: FORM and MATHEMATICA, respectively.}.

The momentum integrals which have to be evaluated are of the following
form after joining the propagators via Feynman parametrization
\begin{equation}
\int \frac{d^d q}{(2 \pi)^d} e^{-i q_+ (\kappa_2 - \kappa_1) }
\frac{P( k_i, x_i | q )}{[q^2 - L]^n},
\end{equation}
where $P(k_i,x_i|q)$ is a polynomial function in $q$, $k_i$ and the Feynman
parameters $x_i$. The divergencies can be evaluated by expanding the
exponential factor in the integrand: $e^{-iq_+ (\kappa_2 - \kappa_1)}
= 1 - iq_+ (\kappa_2 - \kappa_1) + \dots$. Since the denominator depends
on $q^2$ we can average with respect to possible rotations of $q$.
Due to the light-like character of the vector $n$ only the first few terms
survive after integration (we maximally need to expand up to $q_+^3$).
To reduce the result to the conventional form, we have to remove the
terms proportional to $(\kappa_2 - \kappa_1)^m$ integrating by parts
with respect to the Feynman parameters which play now the r\^ole of
fractions in the light-cone position formalism. Typical integration
for a test function $\tau(y,z)$ looks like
\begin{eqnarray}
i (\kappa_2 - \kappa_1) k_{1+}\ \J \ast \tau (y, z)
=\ \J \ast
\left\{
\delta(y) \tau (0 , z)
+
\frac{\partial \tau (y , z)}{\partial y}
-
\delta (1 - y - z) \tau (y , \bar y)
\right\}, \nonumber
\end{eqnarray}
and similarly for terms with $k_{2+}$. In the above equation we
have introduced the following shorthand notation for the integral
\begin{equation}
\J = \int_{0}^{1} dz \int_{0}^{\bar z} dy
e^{-i k_{1+}( \bar y \kappa_1 + y \kappa_2)
-i k_{2+}( \bar z \kappa_2 + z \kappa_1)} .
\end{equation}
When triple and double integration by parts is required, it turns
out that after the first integration all dangerous terms proportional
to $\delta (1 - y - z)$ that might cause a problem vanish identically
in all cases.

For the diagrams which originate from the expansion of the phase
factor there appear in the integrand terms of the type
\begin{equation}
E(q, k_1, y) =
\frac{1}{( q + y k_1 )_+}
\left[
1 - e^{-i (\kappa_2 - \kappa_1) ( y k_1 + q )_+}
\right] ,
\end{equation}
with $q$ being the momentum of integration and $y$ one of the
Feynman parameters. The best way to treat them is to factorize
the $q$-dependence using the following trivial identity
\begin{equation}
E(q, k_1, y)
= \frac{1}{y}
e^{\frac{q_+}{y} \partial_-^{k_1}}
\frac{1 - e^{ - i ( \kappa_2 - \kappa_1 ) y k_{1+} }}{k_{1+}} .
\end{equation}
Expanding as previously the exponential in power series of $q_+^n$,
we can easily perform the final momentum integration since only a
limited number of lowest order terms contribute due to the light-like
character of the vector $n$.

However, this is not the end of the story since some contributions from
the graphs possess apart form desired structure $k_{1+}k_{2+}$ also
$k_{i+}^2$. Therefore, it is necessary to get rid of them by reduction
$k_{i+}^2 \to k_{1+}k_{2+}$. The general structure of these terms in
the sum of all diagrams is of the form (e.g.\ the coefficient in front
of $k_{1+}^2$)
\begin{equation}
\tau_1 (y) + \tau_2 (y) \delta (z) + \tau_3 (y) \delta ( 1 - y - z ).
\end{equation}
Thus, integration by parts reduces this expression to the desired form
up to terms proportional to $\delta (1 - x - y)$. We get
\begin{eqnarray}
\label{reduction}
&&k_{1+}^2\ \J \ast \tau_1 (y) =
k_{1+}^2\ \J \ast
\left\{
\left[ \tau_1 (y) z + \tau_2 (y) \right] \delta (1 - y - z)
- \tau_2 (y) \delta (z)
\right\}
\nonumber\\
&&\qquad\qquad\qquad
+ k_{1+} k_{2+}\ \J \ast
\{
\left[ \tau_1 (y) z + \tau_2 (y) \right] \delta (1 - y - z)
\nonumber\\
&&\qquad\qquad\qquad\qquad\qquad\qquad
- \left[ \tau_1 (0) z + \tau_2 (0) \right] \delta (y)
- \left[ \tau_1'(y) z + \tau_2'(y) \right]
\} .
\end{eqnarray}
Note also that the resulting function
\begin{equation}
k_{1+}^2
\left[
\tau_1 (y) z + \tau_2 (y) + \tau_3 (y)
\right] \delta (1 - y - z)
\end{equation}
can be discard safely because of the Bose symmetry properties of the
string operators it is convoluted with. Similar equations hold for
$k_{2+}^2$.


\begin{figure}[t]
\begin{center}
\vspace{5.0cm}
\hspace{-1cm}
\mbox{
\begin{picture}(0,220)(270,0)
\put(0,-30)                    {
\epsffile{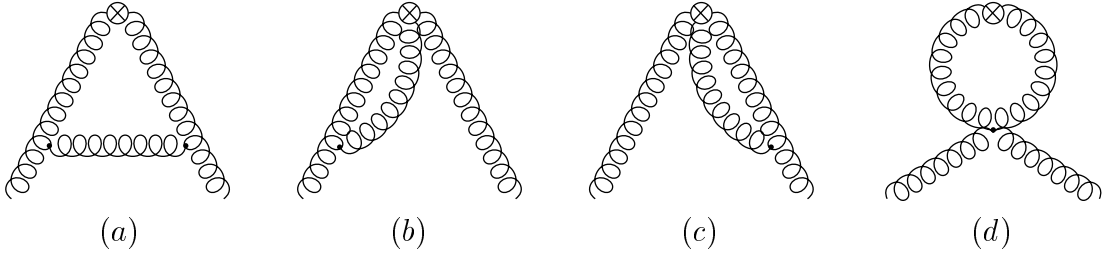}
                               }
\end{picture}
}
\end{center}
\vspace{-9.4cm}
\caption{\label{gluons} One-loop renormalization of the twist-2
gluon operator.}
\end{figure}

Since our result for the ${^{GG}\!\K^-_A}$-kernel is expressed in
terms of the leading order gluonic evolution kernel we find it
useful to give diagram-by-diagram contributions. Namely, straightforward
calculations of the graphs displayed\footnote{Of course, there exists
a diagram analogous to $(d)$ with composite operator instead of the
four-gluon vertex, but it is identically zero in dimensional
regularization since there is no mass parameter entering the
loop integral.} in Fig.\ \ref{gluons} give for parity-odd operators
\begin{eqnarray}
{^{GG}\!\K_{(a)}} (y,z)
\!\!\!&=&\!\!\!
1 - y - z + 4 y z
+ \frac{1}{2} ( 1 - y ) \delta (z)
+ \frac{1}{2} ( 1 - z ) \delta (y)
- ( 1 - \xi ) \frac{1}{2} \delta (y) \delta (z)
\nonumber\\
\!\!\!&+&\!\!\!
\frac{k_{1+}}{k_{2+}}
\left[ 2 y ( 1 - y ) - 1  \right]
+
\frac{k_{2+}}{k_{1+}}
\left[ 2 z ( 1 - z ) - 1 \right] ,
\\
{^{GG}\!\K_{(b+c|NA)}} (y,z)
\!\!\!&=&\!\!\!
- ( 5 + \xi ) \frac{1}{4} \delta (y) \delta (z) \nonumber\\
\!\!\!&+&\!\!\!
\frac{1}{2} \frac{k_{1+}}{k_{2+}}
( 1 - y ) ( 2 - y ) \delta (z)
+
\frac{1}{2} \frac{k_{2+}}{k_{1+}}
( 1 - z ) ( 2 - z ) \delta (y) ,
\\
{^{GG}\!\K_{(b+c|\Phi)}} (y,z)
\!\!\!&=&\!\!\!
\frac{5}{2} \delta (y) \delta (z)
+
\frac{1}{2} \left\{
y - 3 + 2 \left[\frac{1}{y} \right]_+
\right\} \delta (z)
+
\frac{1}{2} \left\{
z - 3 + 2 \left[\frac{1}{z} \right]_+
\right\} \delta (y) ,
\\
{^{GG}\!\K_{(d)}} (y,z)
\!\!\!&=&\!\!\!
- y z \delta ( 1 - y - z ) \nonumber\\
\!\!\!&-&\!\!\! \frac{1}{2}
\frac{k_{1+}}{k_{2+}}
y z \delta ( 1 - y - z )
- \frac{1}{2}
\frac{k_{2+}}{k_{1+}}
y z \delta ( 1 - y - z ) ,
\end{eqnarray}
where the symmetry factors are included into the kernels. Making
the reduction with Eq.\ (\ref{reduction}), where $\tau_1 (y) = 2 y
(1 - y) - 1$ and $\tau_2 (y) = \frac{1}{2} ( 1 - y ) ( 2 - y )$
(and similarly for the $y \to z$ contribution), and summing the
resulting contributions with renormalization constant of the
gluonic fields
\begin{equation}
{^{GG}\!\K_R} (y,z) = - \gamma_G^{(0)} (\xi) \delta (y) \delta (z) ,
\end{equation}
we come to the well known kernel (\ref{GGV-kernel}).

The contributions to the parity even kernel ${^{GG}\!\K^-_A}$
from diagrams in Fig.\ \ref{one-loop} read as follows
\begin{eqnarray}
{^{GG}\!\K^-_{(a+b)}} (y, z)
\!\!\!&=&\!\!\! - 4 i ( \kappa_1 + \kappa_2 )
{^{GG}\!\K}_{(a)} ( y, z | \xi )
+ \W_{(a+b)} (y, z), \\
{^{GG}\!\K^-_{(c)}} (y, z)
\!\!\!&=&\!\!\! - 2 i ( \kappa_1 + \kappa_2 )
{^{GG}\!\K}_{(a)} ( y, z | \xi = 0 )
+ \W_{(c)} (y, z), \\
{^{GG}\!\K^-_{(d+e)}} (y, z)
\!\!\!&=&\!\!\! 2 i ( \kappa_1 + \kappa_2 )
{^{GG}\!\K}_{(b+c)} ( y, z | \xi )
+ \W_{(d+e)} (y, z), \\
{^{GG}\!\K^-_{(f+g)}} (y, z)
\!\!\!&=&\!\!\! - 2 i ( \kappa_1 + \kappa_2 )
\left\{ {^{GG}\!\K}_{(b+c)} ( y, z | \xi )
+ {^{GG}\!\K}_{(b+c)} ( y, z | \xi = 0 ) \right\}
+ \W_{(f+g)} (y, z), \\
{^{GG}\!\K^-_{(h+i)}} (y, z)
\!\!\!&=&\!\!\! 4 i ( \kappa_1 + \kappa_2 )
{^{GG}\!\K}_{(a)} ( y, z | \xi )
+ \W_{(h+i)} (y, z), \\
{^{GG}\!\K^-_{(j+k)}} (y, z)
\!\!\!&=&\!\!\! - 4 i ( \kappa_1 + \kappa_2 )
{^{GG}\!\K}_{(d)} ( y, z )
+ \W_{(j+l)} (y, z), \\
{^{GG}\!\K^-_{(l)}} (y, z)
\!\!\!&=&\!\!\! 2 i ( \kappa_1 + \kappa_2 )
{^{GG}\!\K}_{(d)} ( y, z )
+ \W_{(l)} (y, z).
\end{eqnarray}
From these formulae it is clearly seen how the cancellation of
gauge dependence in the kernels in front of the factor
$( \kappa_1 + \kappa_2 )$ occurs. To be sure in the correctness
of the calculation of the remaining part $\W (y, z)$, we have
kept gauge fixing parameter to be arbitrary. We have observed that
the cancelation of the gauge dependence occurs again between diagrams
with the same structure, similar to the situation for the kernels
which accompany the $( \kappa_1 + \kappa_2 )$-factor. Namely, we have
\begin{eqnarray}
\W_{(a+b+c+h+i)}  (y, z)
&=& \frac{1}{k_{2+}}
\left\{
2 \delta (y) - ( 1 - y ) \delta (z) - 2 y z \delta (1 - y - z)
\right\} \nonumber\\
&+& \frac{1}{k_{1+}}
\left\{
2 \delta (z) - ( 1 - z ) \delta (y) - 2 y z \delta (1- y - z)
\right\}
, \\
\W_{(d+e|NA)}  (y, z)
&=&  \frac{1}{k_{2+}}
\left\{ 1 - 4 y + 2 y^2
- \frac{1}{2} ( 1 + \xi ) \delta (y) \right\} \delta (z) \nonumber\\
&+&
\frac{1}{k_{1+}}
\left\{ 1 - 4 z + 2 z^2
- \frac{1}{2} ( 1 + \xi ) \delta (z) \right\} \delta (y)
, \\
\W_{(d+e|\Phi)}  (y, z)
&=&  \frac{1}{k_{2+}}
\left\{ 2 z - 3 - 2 \left[ \frac{1}{z} \right]_+
+ 2 \left[ \frac{1}{z^2} \right]_+
+ 2 \delta (z) \right\} \delta (y) \nonumber\\
&+& \frac{1}{k_{1+}}
\left\{ 2 y - 3 - 2 \left[ \frac{1}{y} \right]_+
+ 2 \left[ \frac{1}{y^2} \right]_+
+ 2 \delta (y) \right\} \delta (z)
, \\
\W_{(f+g|NA)}  (y, z)
&=&  \frac{1}{k_{2+}}
\left\{ 3 y - 2 y^2
+ \frac{\xi}{2} \delta (y) \right\} \delta (z)
+
\frac{1}{k_{1+}}
\left\{ 3 z - 2 z^2
+ \frac{\xi}{2} \delta (z) \right\} \delta (y)
, \\
\W_{(f+g|\Phi)}  (y, z)
&=&  \frac{1}{k_{2+}}
\left\{ 1 - 2 z + 6 \left[ \frac{1}{z} \right]_+
- 4 \left[ \frac{1}{z^2} \right]_+ \right\} \delta (y)
\nonumber\\
&+& \frac{1}{k_{1+}}
\left\{ 1 - 2 y + 6 \left[ \frac{1}{y} \right]_+
- 4 \left[ \frac{1}{y^2} \right]_+ \right\} \delta (z)
, \\
\W_{(j+k)}  (y, z)
&=&  \frac{1}{k_{2+}}
\left\{ 4 y z - 1 \right\} \delta (1 - y - z)
+ \frac{1}{k_{1+}}
\left\{ 4 y z - 1 \right\} \delta (1 - y - z)
, \\
\W_{(l)}  (y, z)
&=&  \frac{1}{k_{2+}}
\left\{ 1 - 2 y z \right\} \delta (1 - y - z)
+ \frac{1}{k_{1+}}
\left\{ 1 - 2 y z \right\} \delta (1 - y - z)
.
\end{eqnarray}
To simplify the result given by $\W_{(a+b+c+h+i)} (y, z)$, we have
used Eq.\ (\ref{reduction}) and the following identity
\begin{equation}
\J \ast ( k_{1+} + k_{2+} ) \delta (1 - y - z)
= \J \ast \left\{ k_{1+} \delta (z) + k_{2+} \delta (y) \right\}.
\end{equation}
Finally, the kernel which we are interested in is given by
\begin{eqnarray}
{^{GG}\!\K^w} ( y, z ) &=& \W (y, z) + 2
\left(
\frac{1}{k_{1+}} + \frac{1}{k_{2+}}
\right)
\left(
\frac{\beta_0}{2} - \gamma_G^{(0)} (\xi = 0)
\right)
\delta (y) \delta (z) \nonumber\\
&=& \frac{C_A}{k_{2+}}
\left\{
4 \left[ \frac{1}{z} \right]_+
-
2 \left[ \frac{1}{z^2} \right]_+
\right\} \delta (y)
+ \frac{C_A}{k_{1+}}
\left\{
4 \left[ \frac{1}{y} \right]_+
-
2 \left[ \frac{1}{y^2} \right]_+
\right\} \delta (z) ,
\label{GG-even-W}
\end{eqnarray}
where we restored the colour factors omitted previously. This kernel
arises as a factor convoluted with $\J$. Of course, it is easy to
integrate by parts to get rid of momentum factors $k_{i+}$ and
acquire instead $i (\kappa_2 - \kappa_1)$. Adding then the contribution
with $\kappa_2 \leftrightarrow \kappa_1$ gives the resulting kernel in
front of the Bose symmetrical gluon factor (\ref{Abel-gluon}). However,
we will skip this last trivial step since we can obtain the desired
momentum fraction $w$-function from equations similar to
(\ref{GG-even-W}) with the substitution\footnote{Do not mix this $y$
variable (which is the momentum fraction in the ER-BL kernel) with
$y$ in the light cone position formalism which is integrated out
with Fourier transformation.} $k_{1+} = y$, $k_{2+} = \bar y$ and
Fourier transformation.

Finally, let us mention that the results of the calculations for
the parity odd gluonic sector are the same as given above in
Eq.\ (\ref{GG-even-W}). The differences in the intermediate steps
arise from the fact that the contributions of the type
$y z \delta (1 - y - z)$ are absent by simple reason of Bose
symmetry of the light-ray operators which are convoluted with this
kernel.

\setcounter{equation}{0}

\section{Properties of hypergeometric functions.}
\label{hypergeometric}

In this appendix we list all formulae which are necessary for the
evaluation of the Gegenbauer moments of the evolution kernels.

The generalized hypergeometric function is defined as infinite
series \cite{BatErd53}
\begin{equation}
\label{HF-series}
{_m F_n}
\left(\left. { \alpha_1 , \dots , \alpha_m
\atop \beta_1 , \dots , \beta_n }
\right| x \right)
=
\sum_{\ell = 0}^{\infty}
\frac{ (\alpha_1)_\ell \dots (\alpha_m)_\ell }{
(1)_\ell (\beta_1)_\ell \dots (\beta_n)_\ell } x^{\ell},
\end{equation}
where we have introduced the Pochhammer symbol $(\alpha)_\ell =
\Gamma (\alpha + \ell) / \Gamma (\alpha)$. The integral
representation of the hypergeometric function is
(where ${_1F_0} (\alpha | xy) = ( 1 - xy )^{-\alpha}$)
\begin{eqnarray}
\label{B-transformation}
&&\!\!\!\!{_{m + 1}F_{n + 1}}
\left( \left. { \alpha_1 , \dots , \alpha_{m + 1}
\atop \beta_1 , \dots , \beta_{n + 1} }
\right| x \right) \\
&&\!\!\!\!\qquad\qquad=
\frac{\Gamma (\beta_{n + 1})}{
\Gamma (\alpha_{m + 1}) \Gamma (\beta_{n + 1} - \alpha_{m + 1})}
\int_{0}^{1} dy
y^{\alpha_{m + 1} - 1}
\bar y^{\alpha_{m + 1} - \beta_{n + 1} - 1}
{_m F_n}
\left( \left. { \alpha_1 , \dots , \alpha_m
\atop \beta_1 , \dots , \beta_n }
\right| x y \right) . \nonumber
\end{eqnarray}

When one of the upper indices $\alpha_i = - j$, where $j \in {\rm I\!N}$,
the infinite series (\ref{HF-series}) degenerates into a finite polynomial
of order $j$. Therefore, the relation between the Gegenbauer polynomials
and the hypergeometric function reads
\begin{equation}
\label{CtoF}
C_j^\nu (2x - 1) = (- 1)^j
\frac{\Gamma (2\nu + j)}{\Gamma (2 \nu)\Gamma (j + 1)}
{_2F_1}
\left( \left. { - j , j + 2 \nu \atop \nu + \frac{1}{2} }
\right| x \right).
\end{equation}

In the calculation of the integrals we use the following
representation of the Gegenbauer polynomials ($w (x | \nu)
= (x \bar x)^{\nu - \frac{1}{2}}$)
\begin{equation}
\label{C-diff}
\frac{w (x | \nu)}{N_j(\nu)} C_j^\nu (2x - 1)
=
2^{2 \nu} \frac{\Gamma (\nu) (\nu + j)}{
\Gamma (\frac{1}{2})\Gamma (\nu + j + \frac{1}{2})}
(- 1)^j
\frac{d^j}{d x^j}
(x \bar x)^{j + \nu - \frac{1}{2} } .
\end{equation}
Integrating $j$-times by parts we end up with $j$ derivatives acting
on the hypergeometric function. This differentiation can be done with
the help of the formula
\begin{eqnarray}
\label{F-diff}
&&\!\!\!\!\!\!\!\!\!\frac{d^j}{d x^j} x^k
{_m F_n}
\left( \left. { \alpha_1 , \dots , \alpha_m
\atop \beta_1 , \dots , \beta_n }
\right| x \right) \\
&&\!\!\!\!\!\!\!\!\!=
\frac{ (\alpha_1)_{j - k} \dots (\alpha_m)_{j - k} j! }{
(\beta_1)_{j - k} \dots (\beta_n)_{j - k} ( j - k )!}
{_{m + 1} F_{n + 1}}
\left( \left. { \alpha_1 + j - k , \dots , \alpha_m + j - k , j + 1
\atop \beta_1 + j - k , \dots , \beta_n + j - k , j - k + 1  }
\right| x \right) , \ \mbox{for}\ j - k \geq 0 . \nonumber
\end{eqnarray}
To simplify the result of integration it is instructive to use a
relation between the functions of $(m,n)$ and $(m - 1,n - 1)$ orders
\begin{eqnarray}
\label{reduce}
{_m F_n}
\left( \left. { \alpha_1 , \dots , \alpha_m
\atop \beta_1 , \dots , \beta_{n - 1} , \alpha_m + 1 }
\right| x \right)
&=&
\frac{\alpha_m}{1 + \alpha_m - \beta_{n - 1}}
{_{m - 1} F_{n - 1}}
\left( \left. { \alpha_1 , \dots , \alpha_{m - 1}
\atop \beta_1 , \dots , \beta_{n - 1} }
\right| x \right) \\
&-&
\frac{\beta_{n - 1} - 1}{1 + \alpha_m - \beta_{n - 1}}
{_m F_n}
\left( \left. { \alpha_1 , \dots , \alpha_m
\atop \beta_1 , \dots , \beta_{n - 1} - 1 , \alpha_m + 1 }
\right| x \right) . \nonumber
\end{eqnarray}
The number of consequent iterations is governed by the condition
of fulfilling of the relation $1 + \sum_i \alpha_i = \sum_j \beta_j$.
Once being satisfied it allows to apply Saalsch\"utz's theorem
\cite{BatErd53}
\begin{equation}
{_3 F_2}
\left( \left. { - j , j + \alpha , \beta
\atop \gamma , 1 + \alpha + \beta - \gamma }
\right| 1 \right)
=
\frac{( \gamma - \beta )_j ( 1 + \alpha - \gamma )_j}{
( \gamma )_j ( 1 + \alpha + \beta - \gamma )_j} .
\end{equation}
From this relation follows immediately
\begin{equation}
{_3 F_2}
\left( \left. { - j , j + \alpha , \beta
\atop \alpha - 1 , 1 + \beta }
\right| 1 \right)
=
\frac{\beta}{\alpha - 1} \frac{(1)_j}{(\alpha)_j}
\left[
(- 1)^j + \frac{\alpha - \beta - 1}{\beta}
\frac{(\alpha - \beta)_j}{(1 + \beta)_j}
\right] .
\end{equation}
For ${_2 F_1}$ we have
\begin{equation}
{_2 F_1}
\left( \left. { - j , j + \alpha
\atop \beta }
\right| 1 \right)
= (- 1)^j \frac{(1 + \alpha - \beta)_j}{(\beta)_j},
\qquad
{_2 F_1}
\left( \left. { - j + k , j + k + 3
\atop 2 k + 4 }
\right| 1 \right)
= \delta_{jk}.
\end{equation}
Since for the regularized functions we get the derivatives of the
hypergeometric function ${_3 F_2}$ with respect to an index, we
need the following expansion
\begin{equation}
\label{deriv3F2}
{_3 F_2}
\left( \left. { - j , j + \alpha , \beta
\atop 1 + \alpha , \beta + \epsilon }
\right| 1 \right)
= \frac{\Gamma (1 + \alpha)}{\Gamma (1 + \alpha + j)}
\left\{
\delta_{j0}
+
\epsilon \Gamma (j)
\left[
\frac{(1 + \alpha - \beta)_j}{(\beta)_j} - (- 1)^j
\right]
\right\} + {\cal O} (\epsilon^2) ,
\end{equation}
which can be derived making use of a fundamental identity for
${_3 F_2}$-functions \cite{BatErd53} and the definition (\ref{HF-series}).

\end{document}